%% file: planck_high-z_eng.tex
\documentclass{astlb}

\usepackage{graphicx}
\usepackage{color}

\usepackage{apjfonts}

\usepackage{multirow}
\usepackage[hyperfootnotes=false,draft]{hyperref}

\definecolor{darkblue}{rgb}{0,0,0.9}

\def\smfigure#1#2#3{
  \begin{minipage}{1.0\columnwidth}
    \begin{minipage}{0.049\columnwidth}
      \rotatebox{90}{\footnotesize\phantom{0000}#3}
    \end{minipage}
    \begin{minipage}{0.9\columnwidth}
     \includegraphics[viewport=40 188 556 678,width=0.97\columnwidth]{#1}
      \centerline{\footnotesize #2}
    \end{minipage}

    \vskip 3pt
    ~
  \end{minipage}
}

\def\smfiguresmall#1#2#3{
  \begin{minipage}{0.95\columnwidth}
    \begin{minipage}{0.049\columnwidth}
      \rotatebox{90}{\footnotesize\phantom{0000}#3}
    \end{minipage}
    \begin{minipage}{0.95\columnwidth}
      \includegraphics[viewport=40 188 556 540,width=0.97\columnwidth]{#1}
      \centerline{\footnotesize #2}
    \end{minipage}

    \vskip 3pt
    ~
  \end{minipage}
}

\def\doubleline{\vskip 3pt\hrule \vskip 1.5pt \hrule \vskip 5pt}

\begin{document}

\journalinfo{2018}{0}{0}{1}[0]

\title{Optical identifications of high-redshift galaxy clusters from
  Planck Sunyaev-Zeldovich survey}

\author{R.A.~Burenin\email{rodion@hea.iki.rssi.ru}\address{1},
  I.F.~Bikmaev\address{2,3},
  I.M.~Khamitov\address{2,4},
  I.A.~Zaznobin\address{1},\\
  G.A.~Khorunzhev\address{1},
  M.V.~Eselevich\address{5},
  V.L.~Afanasyev\address{6},
  S.N.~Dodonov\address{6},\\
  J.A.~Rubi\~no-Mart\'in\address{7},
  N. Aghanim\address{8},
  R.A.~Sunyaev\address{1,9}
  \addresstext{1}{Space Research Institute RAS, Moscow, Russia}
  \addresstext{2}{Kazan Federal University, Kazan, Russia}
  \addresstext{3}{Academy of Sciences of The Republic of Tatarstan, Kazan, Russia}
  \addresstext{4}{Observatory of Scientific and Technological Research Council of Turkey, Antalya, Turkey}
  \addresstext{5}{Institute of Solar-Terrestrial Physics SB RAS, Irkutsk, Russia}
  \addresstext{6}{The Special Astrophysical Observatory of the Russian Academy of Sciences, Nizhnij Arkhyz, Russia}
  \addresstext{7}{Instituto de Astrof\'isica de Canarias, Tenerife, Spain}
  \addresstext{8}{Institut d\'\ Astrophysique Spatiale, Orsay, France}
  \addresstext{9}{Max Planck Institute for Astrophysics, Garching, Germany}
}

\shortauthor{R. A. Burenin et al.}

\shorttitle{High-redshift galaxy clusters from Planck SZ survey}


\submitted{\today}

\begin{abstract}
  We present the results of optical identifications and spectroscopic
  redshifts measurements for galaxy clusters from 2-nd \emph{Planck}
  catalogue of Sunyaev-Zeldovich sources (\emph{PSZ2}), located at
  high redshifts, $z\approx0.7$--$0.9$.  We used the data of optical
  observations obtained with Russian-Turkish 1.5-m telescope (RTT150),
  Sayan observatory 1.6-m telescope, Calar Alto 3.5-m telescope and
  6-m SAO RAS telescope (Bolshoi Teleskop Alt-azimutalnyi, BTA).
  Spectroscopic redshift measurements were obtained for seven galaxy
  clusters, including one cluster, PSZ2\,G$126.57\!+\!51.61$, from the
  cosmological sample of \emph{PSZ2} catalogue. In central regions of
  two clusters, PSZ2\,G$069.39\!+\!68.05$ and
  PSZ2\,G$087.39\!-\!34.58$, the strong gravitationally lensed
  background galaxies are found, one of them at redshift $z=4.262$.
  The data presented below roughly double the number of known galaxy
  clusters in the second \emph{Planck} catalogue of Sunyaev-Zeldovich
  sources at high redshifts, $z\approx0.8$.

  
\end{abstract}

\section{Introduction}

In \emph{Planck} all-sky survey, by means of the observations of
Sunyaev-Zeldovich effect \citep[SZ,][]{sz72}, the most massive
clusters in observable Universe are detected more or less uniformly
over the entire sky \citep{PSZ1,PSZ2}. This sample is unique and is of
great interest for various cosmological studies such as obtaining the
cosmological parameter constraints using the measurements of the
galaxy cluster mass function
\citep[e.g.][]{av09a,av09b,bv12,PSZcosm13,PSZ2cosm}. In particular,
virtually all the clusters discovered in \emph{Planck} SZ survey will
be included into the large cosmological samples of massive galaxy
clusters that will be selected in the future Spectrum-Roentgen-Gamma
(SRG) / eROSITA all-sky X-ray survey.

The second \emph{Planck} Sunyaev-Zeldovich sources catalogue, released
in 2015 \citep[\emph{PSZ2},][]{PSZ2}, contains 1653 objects of which
1203 are confirmed as massive galaxy clusters. Among the unidentified
SZ sources there is some number of false detections, but also there is
a significant number of unidentified galaxy clusters. Our group take
part in the work on optical identifications and redshift measurements
of the galaxy clusters from this survey
\citep{PSZ1,PSZ_RTT150,PSZ1Addendum,PSZ2,PSZ_Canary,vorobyev16,azt33ik16,br17,amodeo17}.

The completeness of optical identification of Sunyaev-Zeldovich
sources in \emph{Planck} survey changes with redshift. Nearby rich
clusters located at redshifts below $z\approx0.3$ currently are well
studied and, in the vast majority, were known before \emph{Planck}
survey was started. Clusters at higher redshifts, up to $z\approx0.7$,
can be detected in the Sloan Digital Sky Survey (SDSS) images and
also they can be quite easily observed with modern 1.5--3-m
telescopes. Therefore, the completeness of the optical identification
of clusters at such redshifts is high \citep{PSZ1,PSZ1Addendum,PSZ2}.

The optical identification of more distant clusters is generally much
more complicated, so the completeness of the \emph{Planck} catalog
clusters should be lower. In this paper we present the results of our
systematic search for the distant galaxy clusters among the objects
from second catalogue of \emph{Planck} SZ sources, using the data in
optical and near IR. We present the data on the optical identification
and spectroscopic redshift measurements of seven galaxy clusters at
$z>0.7$. These data roughly doubles the number of known galaxy clusters
in the second catalogue of \emph{Planck} SZ sources at these high
redshifts.

\section{Object selection}

Preliminary selection of objects was done using the data of WISE
\citep{wright10}, SDSS \citep{sdssdr13} and Pan-STARRS1 \citep{ps1}
sky surveys. We used the images from WISE survey in 3.4~$\mu$m band,
where all the stars in the field were subtracted using WISE PSF model,
while the stars were identified using SDSS data. WISE images cleaned
from stars were then smoothed using $\beta$-model with $24\arcsec$
radius (which corresponds to the radius of 180~kpc at $z=0.8$). In
these images clusters of galaxies can be identified up to high
redshifts $z\approx1$--$2$, and their IR luminosity correlates well
with cluster mass \citep[e.g.,][]{br15,br17}.

Clusters located at redshifts below $z\approx0.6$ can be identified
using the data of SDSS survey \citep{redmapper14}, using additional
data from WISE all-sky survey it is possible to identify galaxy
clusters at higher redshifts, up to $z\approx0.7$ \citep{br17}. To
identify clusters at even higher redshifts, deeper direct imaging data
in red and near IR bands are required. In our work for these purposes
was used data of the Pan-STARRS1 survey and also the observations at
various optical telescopes (see below).

\begin{figure}
  \centering 
  
  \smfigure{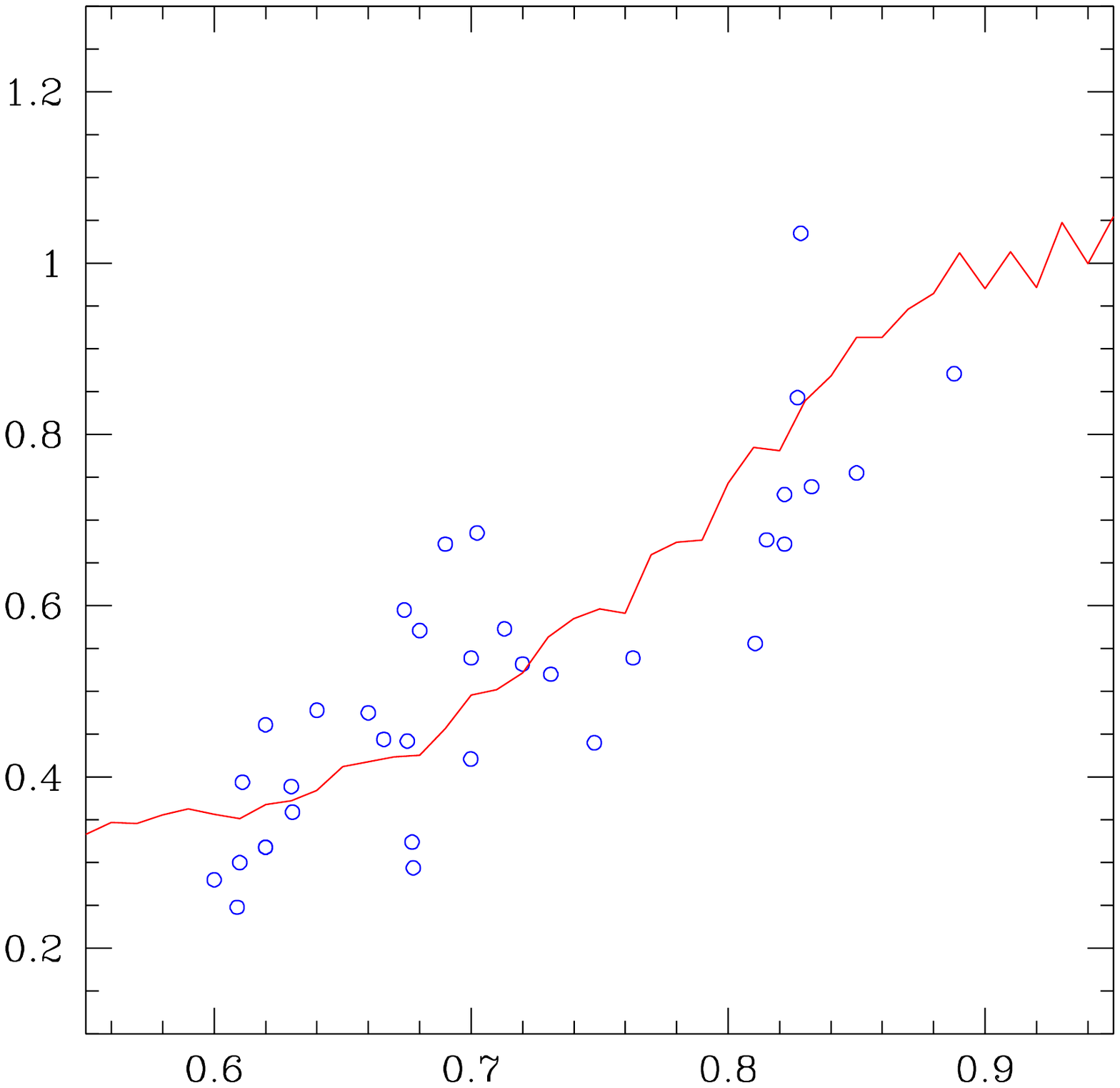}{redshift, $z$}{color, $i-z$}
  
  \caption{The relation between cluster redshifts and median $i-z$
    colors of the brightest cluster galaxies measured using
    Pan-STARRS1 photometric data. The red line shows the color of
    synthetic stellar population spectrum with 11~Gyear age and
    $Z=0.02$ metallicity taken from \cite{bc03}.}
  \label{z_iz}
\end{figure}

It turns out that in Pan-STARRS1 images only the brightest galaxies in
clusters at redshifts $z\approx0.7$--$0.9$ can be detected. However,
from the colors of these galaxies is possible to obtain a rough
estimate of their redshift. In Fig.~\ref{z_iz} the relation between
the median color $i-z$ of the brightest galaxies in clusters and
cluster redshifts is shown. The list of high-redshift galaxy clusters
here are taken from the ROSAT 400 square degree X-ray cluster survey
\citep{400d} and from PSZ2 catalogue. The red line shows the color of
synthetic stellar population spectrum with 11~Gyear age and $Z=0.02$
metallicity taken from \cite{bc03}. One can see that using these data
it is possible to obtain a photometric redshift estimate with the
accuracy not worse than approximately $\delta z/(1+z)\approx 0.03$, in
$0.7<z<0.9$ redshift range.

Using these data, 18 objects from PSZ2 catalogue were selected. Three
of them were discarded on the basis of deep direct images of these
fields (see below), observations of eight objects will be continued.
Seven Sunyaev-Zeldovich sources were identified with high-redshift
galaxy clusters, located at redshifts above $z\approx0.7$. For the
brightest galaxies in these clusters optical spectra were obtained,
which allowed to measure cluster spectroscopic redshifts. These the
observations are discussed in detail below.

\input{fig_images_eng.tex}

\input{fig_spectra_eng.tex}

\begin{figure*}
  \centering
  \vskip -0.3cm
  \smfiguresmall{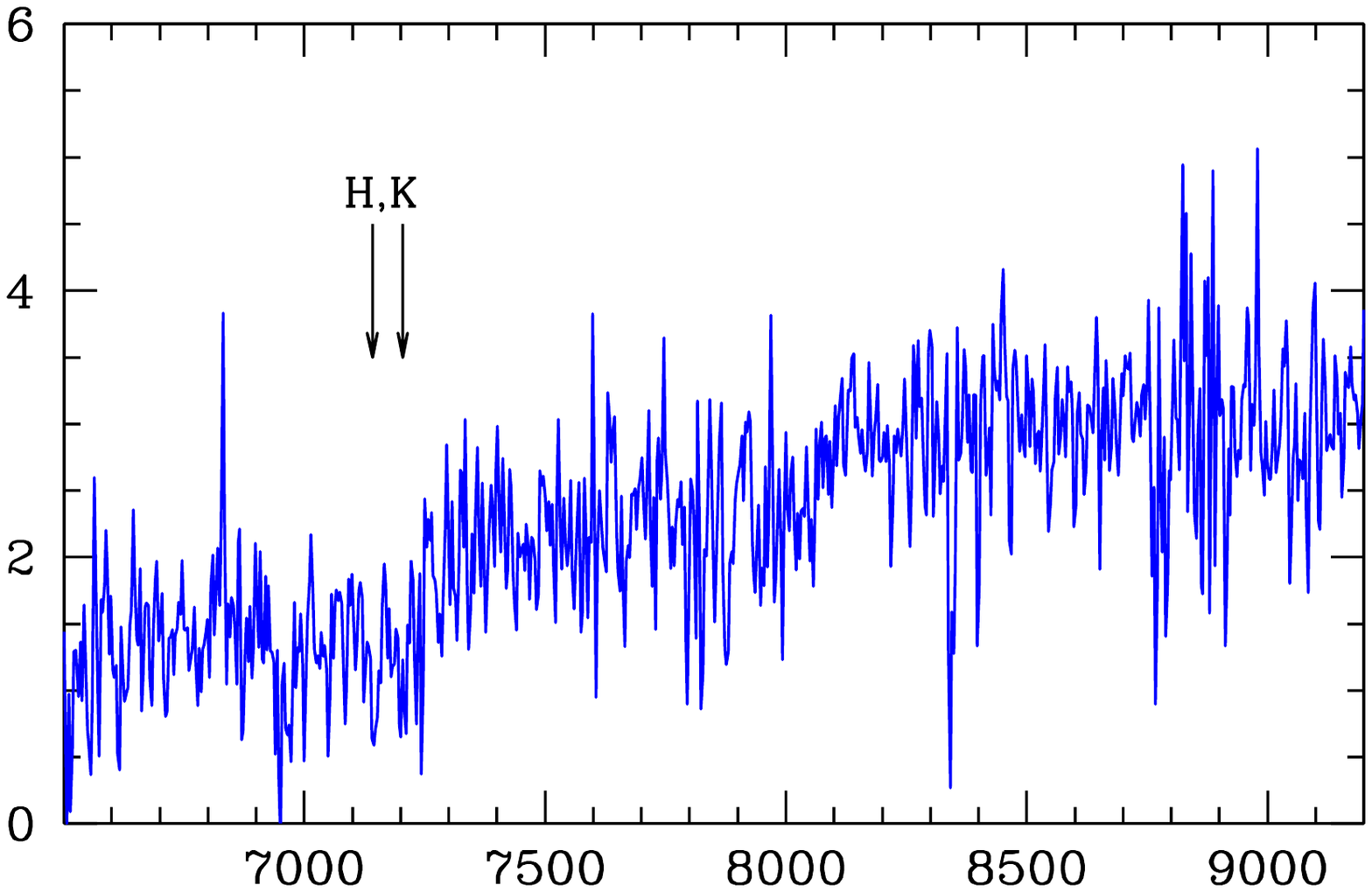}{$\lambda$, \AA}{Flux, $10^{-17}$~erg\,s$^{-1}$\,cm$^{-2}$}
  ~
  \smfiguresmall{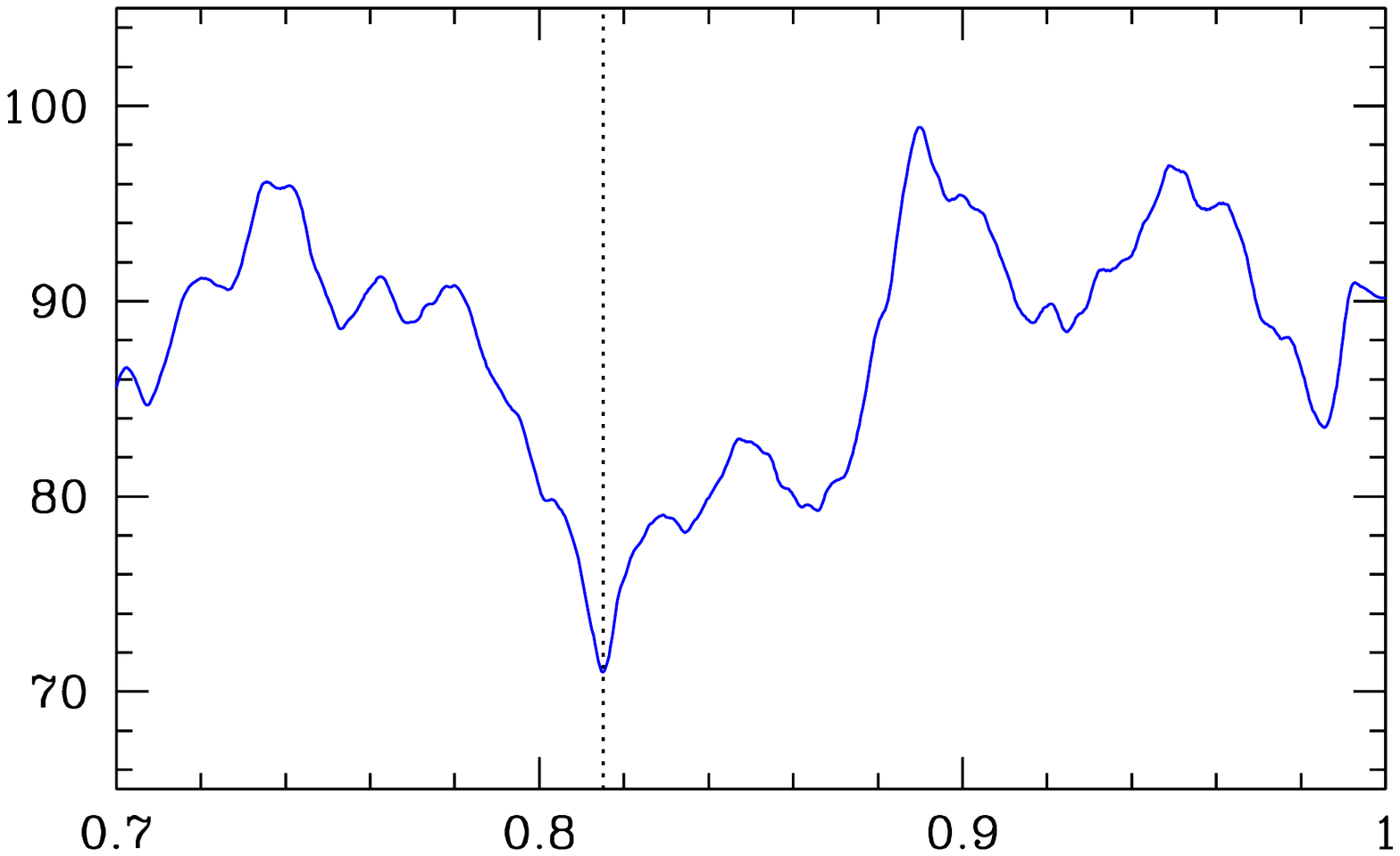}{$z$}{$\chi^2$} 

  \caption{Spectroscopic measurement of PSZ2\,G$126.57\!+\!51.61$
    galaxy cluster redshift. Left panel: spectrum of brightest cluster
    galaxy obtained at Sayan observatory 1.6-m telescope using ADAM
    spectrograph with 6~h total exposure. Right panel: the value of
    $\chi^2$ obtained in result of cross correlation of this spectrum
    with the spectral template of elliptical galaxy.}
  \label{fig:spec_G12657}
\end{figure*}

\section{Observations}

Imaging and spectroscopic observations of objects selected as
high-redshift, $z>0.7$, galaxy cluster candidates were made at various
telescopes available to our group.  Deep direct images in SDSS filters
\textit{ri} were obtained with 1.5-m Russian-Turkish telescope
(RTT150) using the
\textit{TFOSC}\footnote{http://hea.iki.rssi.ru/rtt150/en/index.php?page=tfosc}
instrument, with 6-m telescope of SAO RAS (Bolshoi Telescop
Alt-azimuthalny, BTA-6m) using SCORPIO and SCORPIO-2 instruments
\citep{scorpio05,scorpio11}, and also at the Calar Alto 3.5-m
telescope using
\textit{MOSCA}\footnote{http://w3.caha.es/CAHA/Instruments/MOSCA/}
instrument. In some cases when SDSS \textit{ri} filters were not
available, Bessel \textit{RI} filters were used.

The exposures were chosen to reach the magnitude limits not worse than
about $25^m$ in \textit{r} and $24^m$ in \textit{i} bands
($5\sigma$). This required the 1--2~h exposures at 1.5-m telescope
RTT150, 15--30~min exposures at Calar Alto 3.5-m telescope and
5--10~min at BTA 6-m telescope. All of these the data were obtained at
seeing not worse than $1.5\arcsec$. The total exposure was split into
few smaller exposures so that in each image the number of background
counts does not exceed a few thousands ADU.  Between the exposures the
telescope pointing was shifted by 10--20\arcsec. During the data
reduction the images were corrected for bias and flat field, the
images in \textit{i,I} bands were further corrected for fringes. The
images were then aligned and combined using the algorithms allowing to
remove cosmic rays and to reduce systematic errors of flat field
calibrations.

The spectra of brightest cluster galaxies were obtained mainly at BTA
6-m telescope using SCORPIO and SCORPIO-2 instruments
\citep{scorpio05,scorpio11}. The spectrum of one object, the brightest
galaxy in PSZ2\,G$126.57\!+\!51.61$ (see below), was obtained at Sayan
observatory 1.6-m telescope using the low and medium resolution
spectrograph ADAM \citep{adam16,azt33ik16}. In these observations the
long slit position angle was set to obtain spectra of at least two red
sequence galaxies in cluster, when possible. The 1--2~h total exposure
was used at BTA 6-m telescope, and 6~h --- at Sayan observatory 1.6-m
telescope. The total exposure was split into 900--1200~s exposures,
between the exposures the objects were shifted for 10--20\arcsec\
along the slit.

The 2d spectra were bias and flat field corrected and transformed to
single dispersion solution along the slit. The flat field and
dispersion solution were calibrated using the spectra of calibration
lamps obtained during every observation. The 2d spectra obtained with
SCORPIO instrument at BTA 6-m telescope were also corrected for
fringes. For SCORPIO-2 instrument at BTA 6-m telescope and for ADAM
spectrograph at Sayan observatory 1.6-m telescope these corrections
are not necessary because these instruments are equipped with thicker
deep depletion CCDs. After that 2d spectra were aligned and combined,
the cosmic rays were eliminated from the resulting 2d spectrum. One
dimensional object and background spectra were extracted in a standard
way. Flux calibration of these spectra was done using the observations
of stars from European Southern Observatory list of spectrophotometric
standards. If needed, the spectra were divided by the normalized
spectra of standard stars to correct for the atmospheric extinction
lines.

All optical data reduction was done using
\emph{IRAF}\footnote{http://iraf.noao.edu/} package, and also using
our own software. In total, for these observations we used about 20~h
of clear dark observing time at RTT150 telescope, about 9~h --- at
Sayan observatory 1.6-m telescope, about 3.6~h --- at Calar Alto 3.5-m
telescope and about 14.5~h --- at BTA 6-m telescope. Note, that much
more observations of clusters from PSZ2 catalogue at lower redshifts
were done at these telescopes. The results of these observations will
be published in subsequent papers.

\section{The results of Observations}

In Fig.~\ref{fig:fcharts} the images from WISE survey in 3.4~$\mu$m
band are shown for every cluster (in the left panels).  These images
are cleaned from stars and convolved with $\beta$-model of $24\arcsec$
radius, which approximately corresponds to 180~kpc linear size at
redshift $z=0.8$. The right panels show the pseudocolor images of the
clusters fields in SDSS \emph{irg} filters which correspond to
\emph{RGB} colors in these images. Images in filters \textit{ri} were
taken from our observations (for some clusters we used the images in
Bessel \textit{RI} filters instead). The images in \textit{g} filter
were taken from SDSS survey and (or) Pan-STARRS1 survey. The centers
of the images correspond to the clusters optical centers, the image
size is $5\arcmin \times 5\arcmin$.

In the centers of the IR images shown in Fig.~\ref{fig:fcharts}, the
IR surface brightness excesses are observed, which indicates the
presence of a large number of galaxies in this region of the sky. Note
that for most objects discussed in this paper, i.e.\ for all objects
except PSZ2\,G$092.69\!+\!59.92$ and PSZ2\,G$237.68\!+\!57.83$, the
presence of increased surface density of the number of galaxies
according to the data of WISE survey was mentioned in the
\textit{PSZ2} catalogue. In pseudocolor images in
Fig.~\ref{fig:fcharts} one can see that these galaxies are weak in the
optical range and all of them are of about the same red color, i.e.,
they constitute the red sequence of the cluster. Using the colors of
these galaxies, including data on color $i-z$ (see above), it is
possible to obtain photometric redshift estimates, which appears to be
higher than $z_{\mbox{\scriptsize phot}}= 0.7$ for all objects.
Spectroscopic redshifts were obtained for the brightest galaxies
within the red sequences of the clusters.

In Fig.~\ref{fig:spec} the examples of spectroscopic measurements of
the cluster redshifts are shown. The left panel shows the measured
spectrum of the brightest cluster galaxies with some spectral features
indicated, right panel --- the $\chi^2$ value obtained in result of
comparison of this spectrum with a template spectrum of an elliptical
galaxy. The redshift measurement error is about
$\delta z\approx0.001$.

The redshift of galaxy cluster PSZ2\,G$126.57\!+\!51.61$ was measured
at Sayan observatory 1.6-m telescope using ADAM spectrograph.  The
brightest galaxy of this cluster is of magnitude $r^\prime=20.3$ and
$i^\prime=18.9$, therefore to measure the redshift of this galaxy at
small 1.6-m telescope the large exposure times were needed.  We used
the spectrum with 6~h total exposure, which was accumulated during the
observations in Spring and Autumn 2017. The resulting spectrum and
$\chi^2(z)$--redshift dependence are shown in
Fig.~\ref{fig:spec_G12657}. In this spectrum it is difficult to detect
individual spectral lines, however, the 4000\AA\ break located near
the calcium \textit{H} and \textit{K} lines is clearly visible, which
allows to obtain a reliable measurement of the cluster redshift,
$z=0.815$.

The optical identifications and spectroscopic redshift measurements
for two clusters, PSZ2\,G$087.39\!-\!34.58$ and
PSZ2\,G$126.28\!+\!65.62$, were obtained recently also at 10-m Gran
Telescopio Canarias (GTC). The results of these measurements will be
published later in separate paper. They are consistent with the
results presented in this paper, but are of higher accuracy since the
redshifts of larger number of galaxies are measured.

Four clusters PSZ2\,G$069.39\!+\!68.05$, PSZ2\,G$087.39\!-\!34.58$,
PSZ2\,G$092.69\!+\!59.92$ and PSZ2\,G$126.28\!+\!65.62$, were
identified as distant galaxy clusters in the extended \emph{Planck} SZ
galaxy cluster catalogue \citep[][B17]{br17}. In this catalogue the
redshifts for the clusters PSZ2\,G$069.39\!+\!68.05$ and
PSZ2\,G$087.39\!-\!34.58$ were taken from SDSS survey. Although the
redshifts of single galaxies at the side from the optical center of
the cluster were used, these redshifts are confirmed by our
measurements presented in this paper. In addition, in case of cluster
PSZ2\,G$087.39\!-\!34.58$, in the latest (14th) SDSS data release the
redshift of central cluster galaxy was published, which agrees well
with our measurement (see below). The cluster
PSZ2\,G$092.69\!+\!59.92$, apparently is a projection of two clusters,
and the redshift of distant cluster, $z=0.848$, is also confirmed by
the latest SDSS data.  For cluster PSZ2\,G$126.28\!+\!65.62$ the
photometric redshift estimate was given in B17, which is also
consistent with our measurements.  The other clusters were not
included in the B17 catalogue, because clusters with photometric
redshift estimates above $z=0.7$ were not considered in B17.

The list of clusters that have been identified as a result of our work
is presented in Table~\ref{tab:clres}. The Table contains the
coordinates of clusters optical centers, spectroscopic redshift
measurements, the number of spectroscopically observed galaxies used
to measure cluster redshift, also some notes on individual clusters
are given.

\input{table_1_eng.tex}

\subsection{Notes on individual objects}

\paragraph{PSZ2\,G$069.39\!+\!68.05$}

\begin{figure*}
  \centering
  
  \includegraphics[width=0.95\columnwidth]{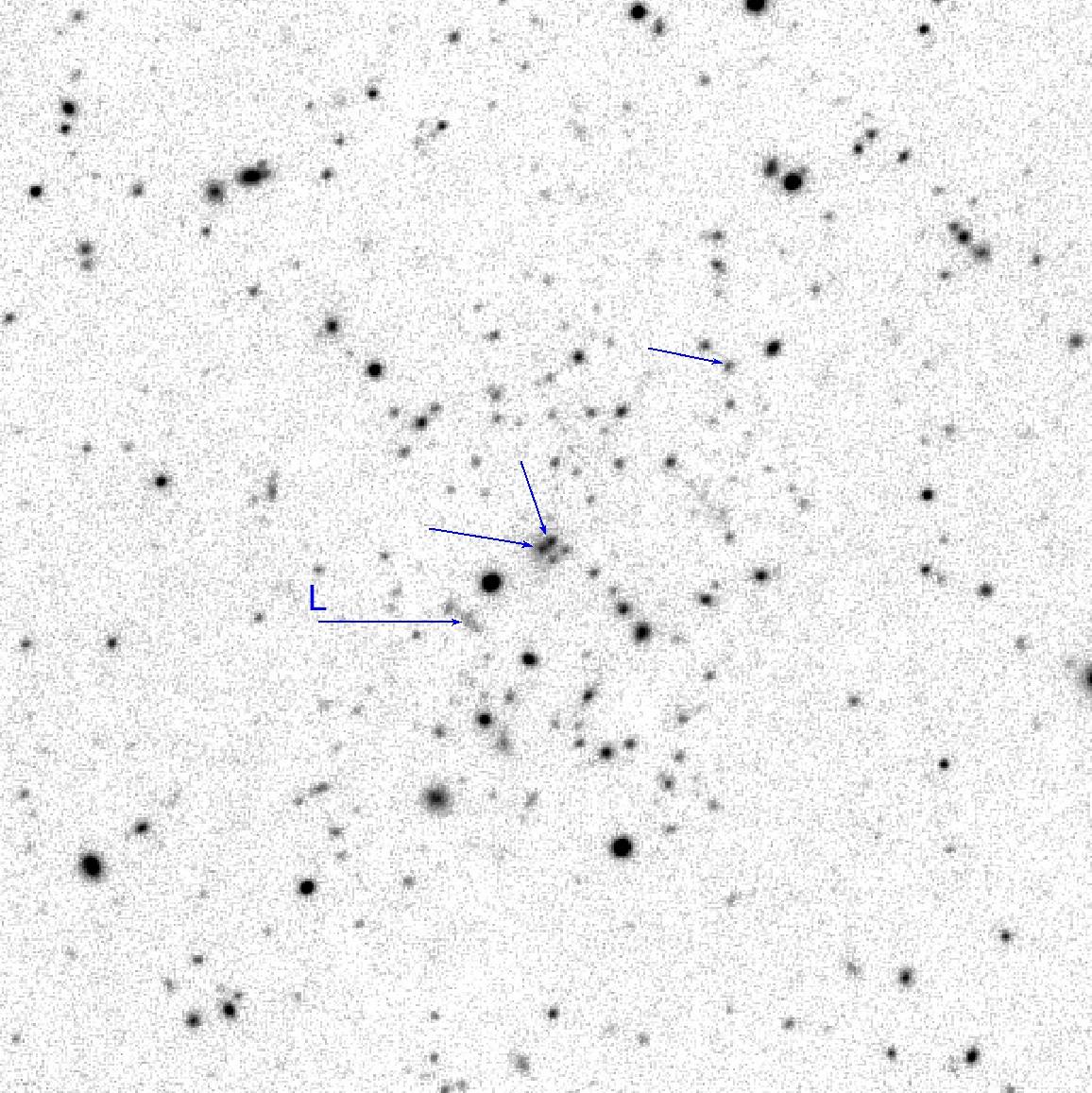}
  ~~
  \begin{minipage}{1\columnwidth}
    \vskip -6cm
    \smfigure{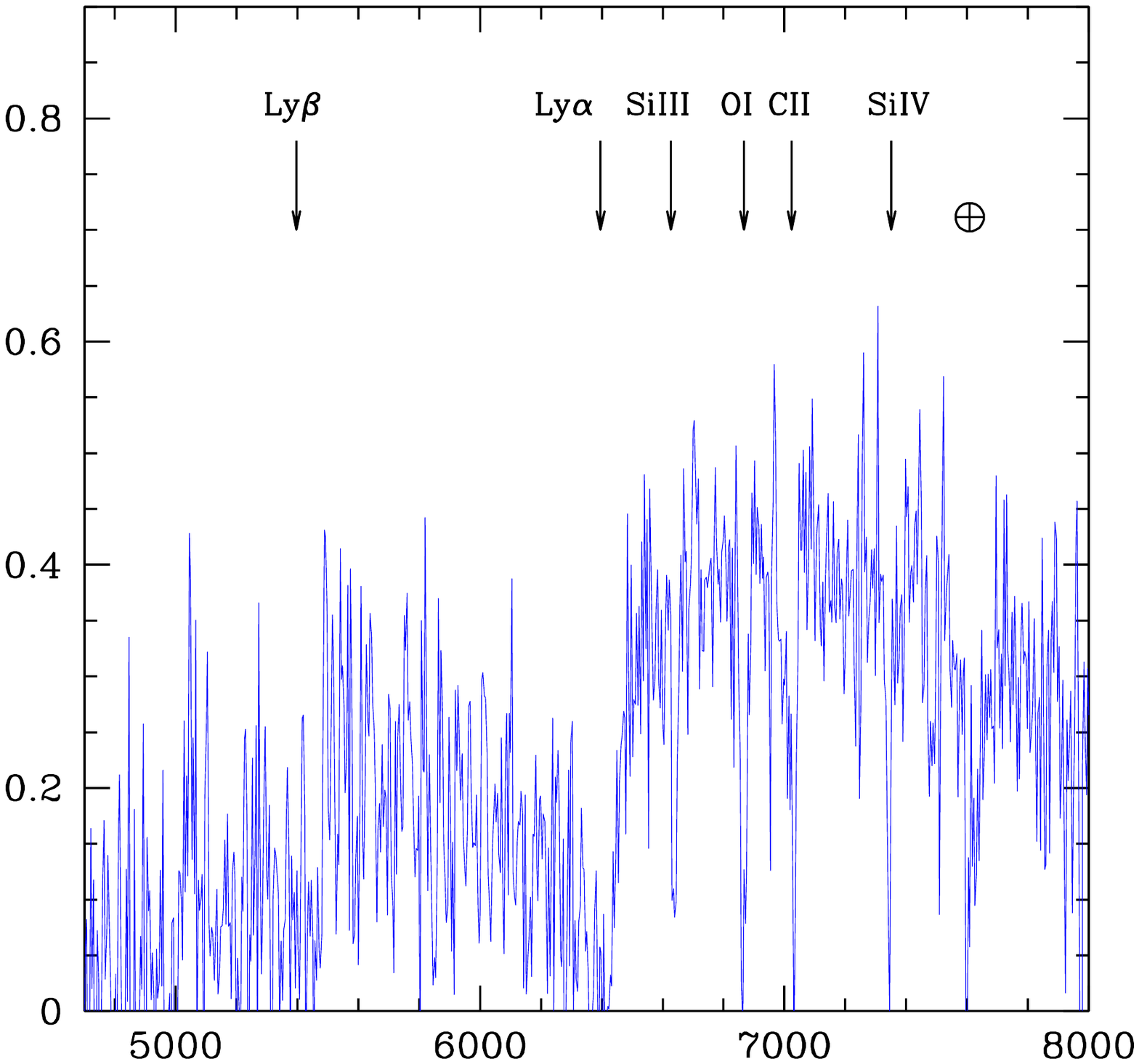}{$\lambda$, \AA}{Flux, $10^{-17}$~erg\,s$^{-1}$\,cm$^{-2}$}
  \end{minipage}
  
  \caption{Left panel: the image of the central part
    ($3\arcmin \times 3\arcmin$) of the cluster
    PSZ2\,G$069.39\!+\!68.05$ field in \emph{I} band, obtained with
    BTA 6-m telescope using SCORPIO-2 instrument with 150~s
    exposure. The objects with measured spectra are shown by
    arrows. The gravitational lensing arc is indicated by the letter
    \emph{L}. Right panel: the spectrum of this arc, obtained at BTA
    6-m telescope using SCORPIO-2 instrument with 3600~s
    exposure. Some spectral features observed at redshift $z=4.262$
    are indicated.}
  \label{fig:G069_lens}
\end{figure*}

In the optical images of the central region of this cluster the
gravitational lensing arc of $7\arcsec$ length was found (see
Fig.~\ref{fig:G069_lens}, left panel). In the spectrum of this galaxy
that was obtained at the 6-m telescope BTA using SCORPIO-2 instrument
with 3600~s exposure (Fig.~\ref{fig:G069_lens}, right panel) the
absorption Lyman series hydrogen lines, as well as narrow absorption
lines of ionized silicon, oxygen and carbon are detected. The redshift
of the galaxy, measured from narrow absorption lines, is
$z=4.262$. Note that in these observations the slit of the
spectrograph was located across the direction of the lens, so only
about 30\% of light from lensed galaxy was observed inside the
slit. The flux integrated over the entire area of the lens corresponds
to stellar magnitude about $20.7$ in \textit{I} band.

In SDSS survey data there is a spectrum of the galaxy located
approximately $2\arcmin$ to the North from the optical center of the
cluster ($\alpha,\delta = $~14:21:39.9, +38:23:07, J2000). In 13th
SDSS data release this spectrum was identified as a spectrum galaxy at
$z=0.7617$, and in 14th data release --- as a spectrum of the quasar
at $z=4.230$. Given that this object is extended and its color is
consistent with the color of cluster red sequence, the first
measurement is likely the correct one. It is consistent with out
spectroscopic redshift measured from the spectra of galaxies indicated
by arrows in Fig.~\ref{fig:G069_lens}. It is this SDSS redshift
measurement was used earlier in the identification of this cluster
presented in B17 catalogue.

\begin{figure}
  \centering
  
  \includegraphics[width=0.95\columnwidth]{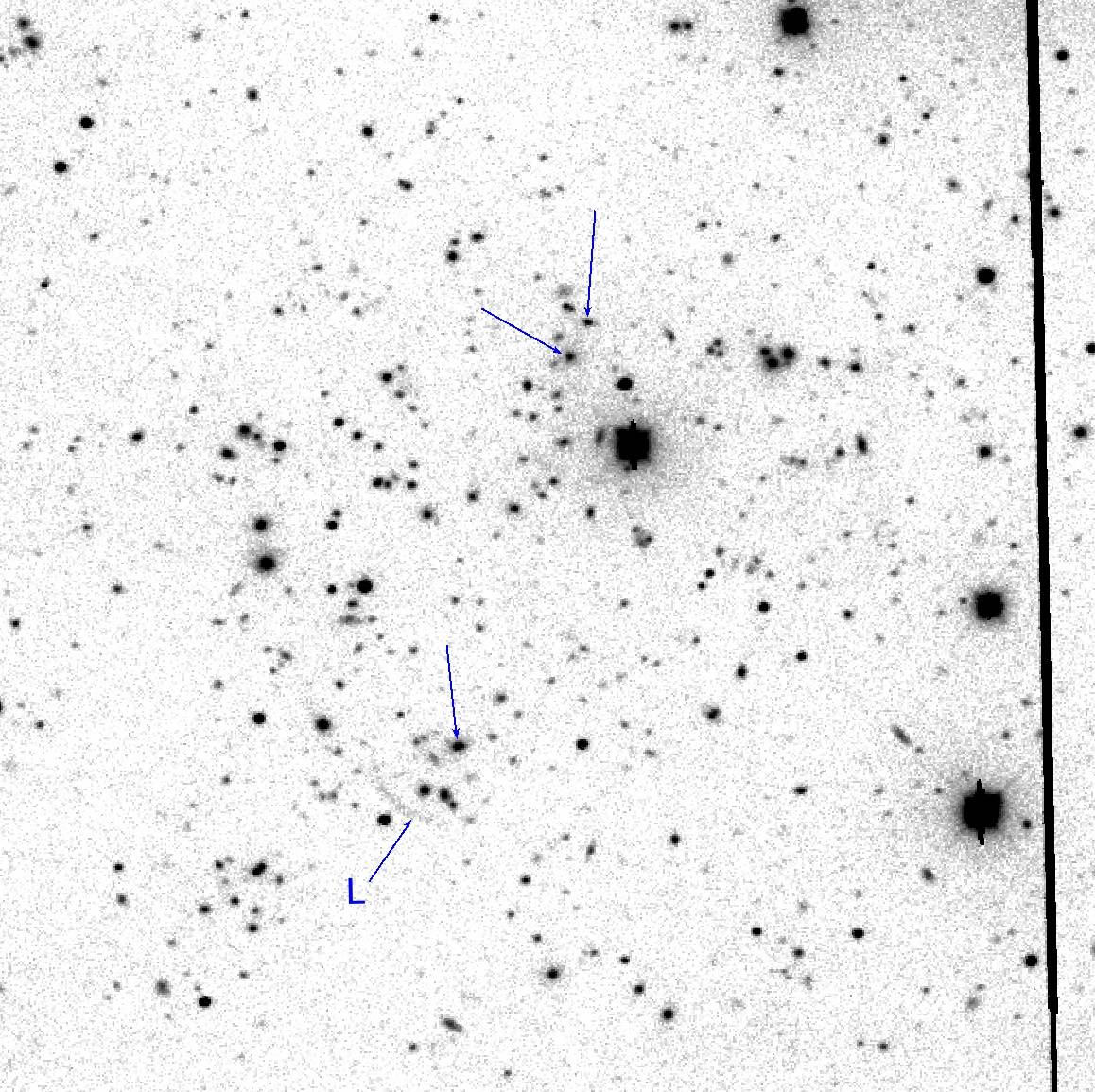}
  
  \caption{The image of the central part ($3\arcmin \times 3\arcmin$)
    of cluster PSZ2\,G$087.39\!-\!34.58$ field in \emph{i} band,
    obtained at Calar Alto 3.5-m telescope using \emph{MOSCA}
    instrument with 900~s exposure. The galaxies, whose spectra were
    used to measure cluster redshift, are shown with arrows. The
    gravitational lensing arc is indicated by letter \emph{L}.}
  \label{fig:G087_spec_obj}
\end{figure}

\begin{figure}
  \centering
  
  \includegraphics[width=0.95\columnwidth]{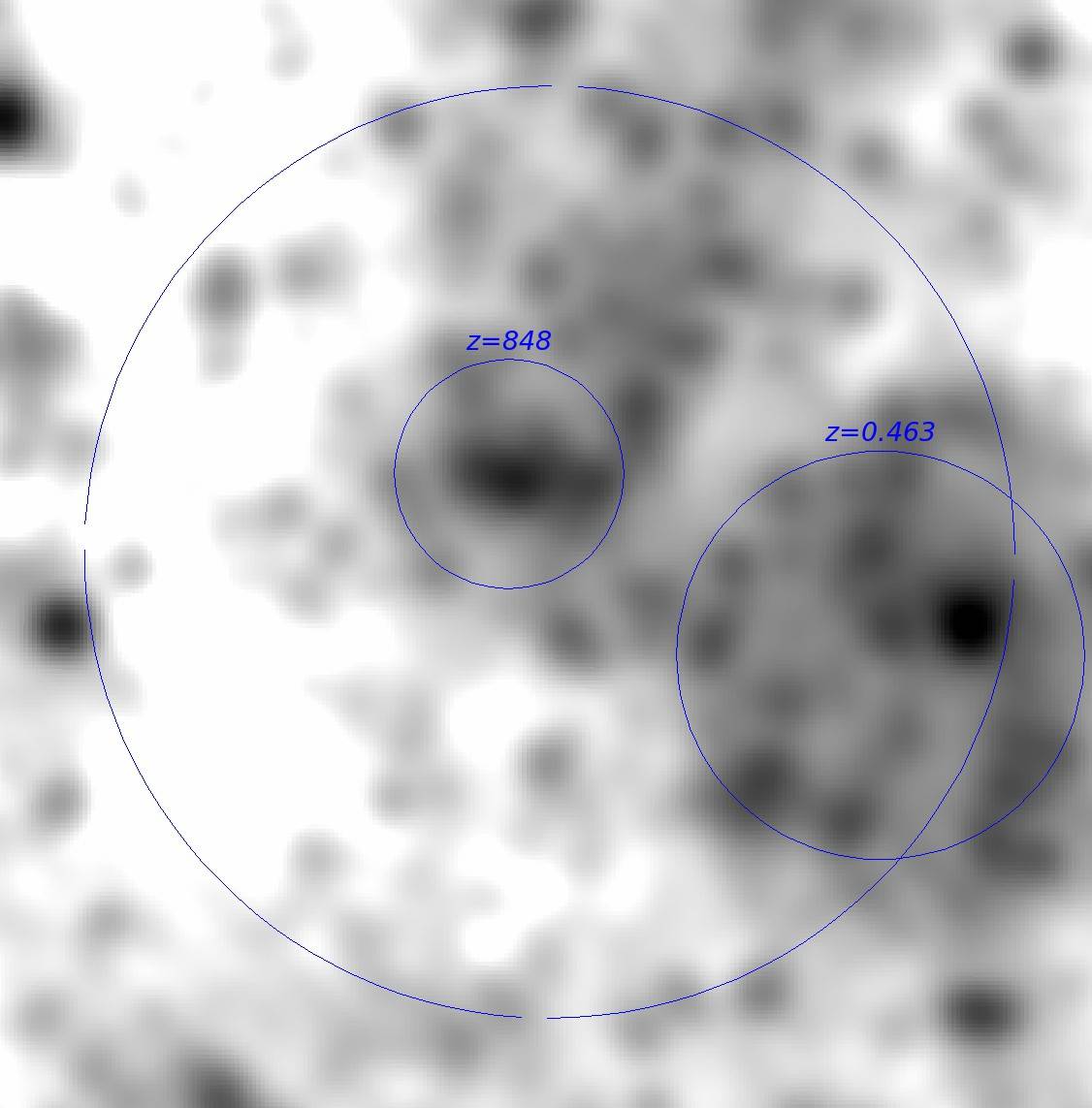}
  
  \caption{The IR image of the field of cluster
    PSZ2\,G$092.69\!+\!59.92$ from WISE survey, cleaned from stars and
    convolved with $\beta$-model of $24\arcsec$ radius. The
    localization of PSZ2 source is shown with a large circle
    ($5\arcmin$ radius). The circles of smaller radii show the
    clusters that are found in this field.}

  \label{fig:G092_wise}
\end{figure}

\paragraph{PSZ2\,G$087.39\!-\!34.58$} 

In the central part of the cluster the gravitational lensing arc of
$14\arcsec$ length was found (see\ Fig.~\ref{fig:G087_spec_obj}).  The
flux of lensed galaxy integrated over the entire area of the lens
corresponds to the \textit{i} band magnitude about $21.7$.

The galaxies, whose spectra were obtained at 6-m BTA telescope with
SCORPIO spectrograph and then used to measure cluster redshift, are
shown with arrows in Fig.~\ref{fig:G087_spec_obj}. Note that in the
latest (14th) SDSS data release there are two measurements of the
redshifts for galaxies in this clusters that are consistent with our
measurements. In B17 catalogue the redshift for this cluster redshift
was measured using only one spectroscopic measurement from SDSS survey
data available at that time. In Table~\ref{tab:clres} much more
reliable redshift measurement for this cluster is provided, which was
obtained using all the available data.

\paragraph{PSZ2\,G$092.69\!+\!59.92$} 

This Sunyaev-Zeldovich source can be identified with the projection
of two clusters of galaxies. One of them, located at redshift
$z=0.463$, was discovered in the \textit{redMaPPer} cluster survey
\citep{redmapper14}. In the B17 catalogue this SZ source was
automatically identified with that cluster because the clusters
redshifts above $z\approx0.7$ were not considered in this
work. However, in the field of this SZ source there is also one more
galaxy cluster, located at redshift $z=0.848$, the image of this
cluster is shown in Fig.~\ref{fig:fcharts}. Both clusters are clearly
visible in the WISE image, which is shown in
Fig.~\ref{fig:G092_wise}. The cluster at $z=0.848$ is found to be
closer to the center of the SZ source and is of much higher IR
luminosity. Therefore, most likely, it is this cluster is most massive
one and provide the main contribution to the Sunyaev-Zeldovich signal
of this \emph{Planck} SZ source.

Note that for the galaxies of both clusters there are spectroscopic
data in SDSS. For more distant cluster there is one spectroscopic
measurement ($z=0.8421$) in SDSS for the galaxy located at the
approximately $3\arcmin$ distance from the optical canter of the
cluster in \emph{NW} direction. This measurement is in good agreement
with our results.

\paragraph{PSZ2\,G$237.68\!+\!57.83$}

\begin{figure}
  \centering
  
  \includegraphics[width=0.95\columnwidth]{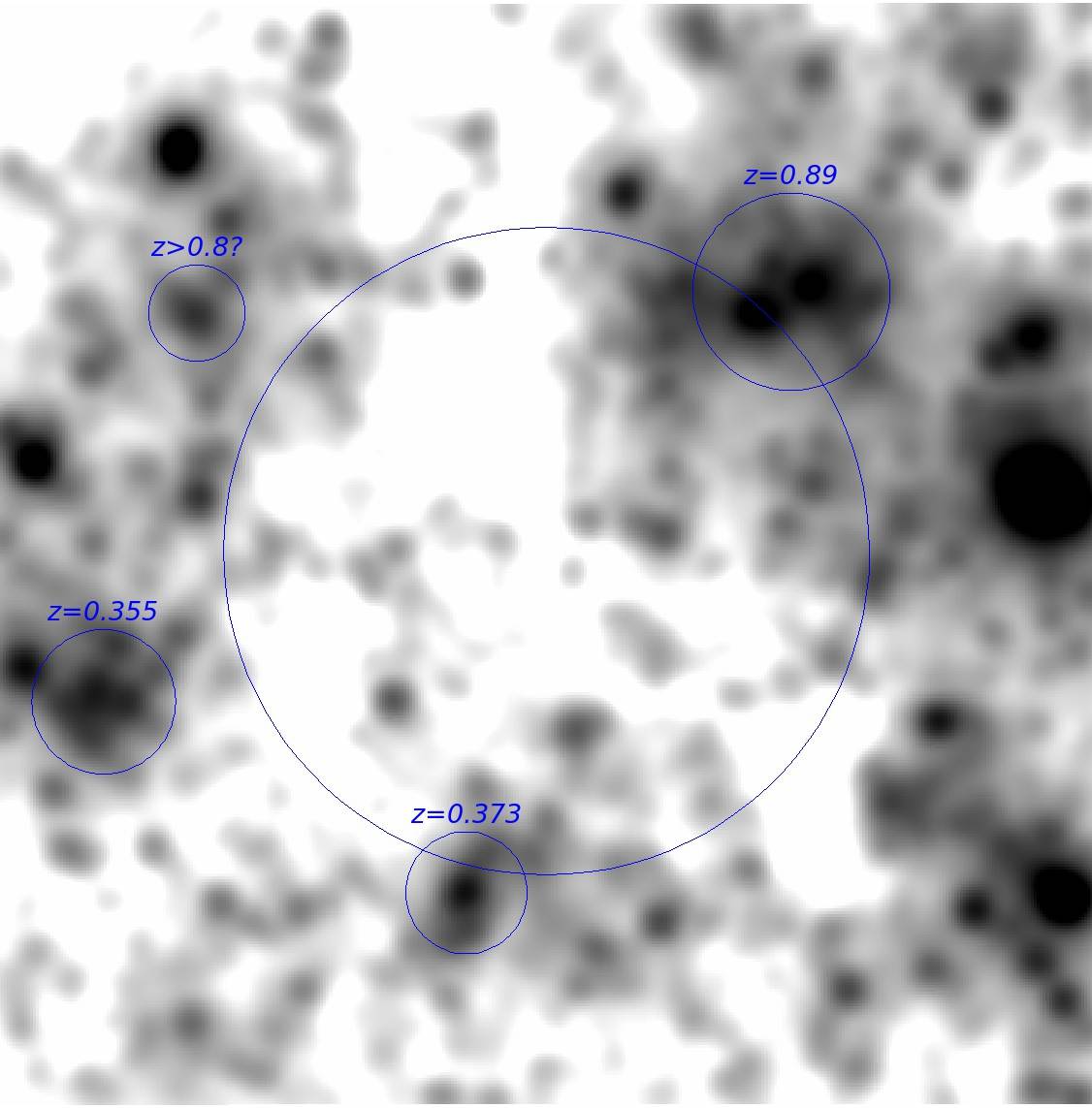}
  
  \caption{The WISE image of the field of PSZ2\,G$237.68\!+\!57.83$ SZ
    source, cleaned from stars and convolved with $\beta$-model of
    $24\arcsec$ radius. The localization of PSZ2 source is shown with
    large circle ($5\arcmin$ radius). The circles of smaller radii
    show the clusters which are found in this field. The remaining
    bright IR sources are connected with nearby galaxies in the
    field.}
  \vskip -0.3cm
  
  \label{fig:G237_wise}
\end{figure}

In the field of this Sunyaev-Zeldovich source it is possible to detect
several clusters of galaxies, as shown in
Fig.~\ref{fig:G237_wise}. Redshifts for clusters at $z=0.373$ and
$z=0.355$ taken from spectroscopic SDSS data.  The cluster at
$z=0.892$, apparently, is of the highest IR luminosity and therefore
should be the most massive one and should give the largest
contribution to the total Sunyaev-Zeldovich signal. However, other
clusters may also give a significant contribution to the observed SZ
signal. In order to measure the contribution of each cluster
additional data in X-rays and (or) mm-band are required.

\begin{figure}
  \centering
  \smfigure{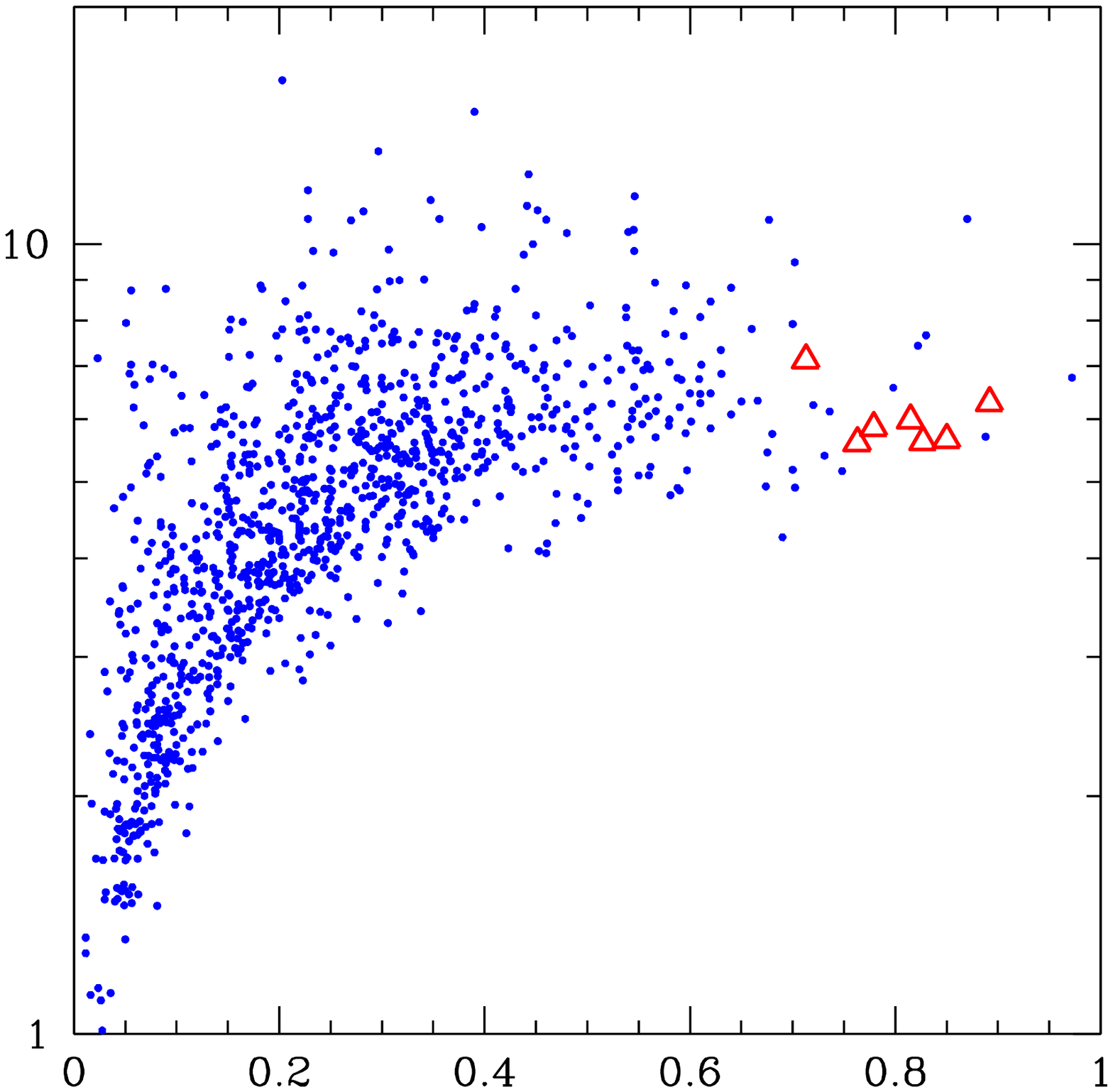}{$z$}{$M_{500}, 10^{14} M_\odot$}

  \caption{The relations between redshifts and masses for clusters
    optically identified in \emph{PSZ2} catalogue (blue points) and in
    our work (red triangles).}
  
  \label{fig:z_msz}
\end{figure}

\section{Conclusions}

At the redshifts above $z\approx 0.5$ in 2-nd \emph{Planck} catalogue
of Sunyaev-Zeldovich sources only the most massive clusters with
masses above $M_{500}\approx 5\cdot 10^{14} M_\odot$ are
detected. These objects are extremely rare at high redshifts. For
example, \emph{PSZ2} catalogue contains 12 clusters of galaxies at
redshifts $z>0.7$, and 5 clusters --- at redshifts $z>0.8$.  In this
paper we present the optical identifications and spectroscopic
redshift measurements for 7 clusters of galaxies at redshifts $z>0.7$,
of them 4 are found at redshifts $z>0.8$. The place of these clusters
at the redshift--mass diagram is shown in Fig.~\ref{fig:z_msz}. Here
we use a simple $M_{500}$ mass estimate made using the \emph{Planck}
$y$-parameter map \citep{PYmap}.  Thus, one can see that the data
presented in our work approximately double the number of known massive
clusters in \emph{PSZ2} catalogue at high redshifts, $z\approx0.8$.

In our list there are few more cluster candidates, some of them may be
identified as distant clusters, located at redshifts $z\approx0.8$ and
higher in result of future observations. Observations of these objects
are continued, however, we can already now state that in result of our
work the completeness of optical identification of SZ sources from
\textit{PSZ2} catalogue for distant galaxy clusters in the Northern
sky, at declination above $\delta>-20^\circ$, increased
significantly. However, some number of distant clusters among the
objects in \textit{PSZ2} catalogue should remain to be
unidentified. In general, we can assume that among the
Sunyaev-Zeldovich sources from \textit{PSZ2} catalogue there should be
of order of 10 more unidentified clusters of galaxies at redshifts
$z>0.7$. Therefore, some incompleteness of the \emph{PSZ2} catalogue
for such clusters should remain.

We note that this incompleteness can not affect the cosmological
results obtained with \emph{Planck} survey galaxy cluster data. The
reason is that the cosmological sample of \emph{PSZ2} catalogue is
constructed using the clusters with higher SZ detection significance
and these clusters are of even higher mass on average. The number of
these even more massive clusters at high redshifts is very small. So,
in cosmological sample of \emph{PSZ2} catalogue there are only three
clusters at $z>0.7$. In our work we identified only one more such
cluster, PSZ2\,G$126.57\!+\!51.61$.

Our group continues the observations of galaxy clusters from the
\emph{Planck} all-sky Sunyaev-Zeldovich survey. In addition to the
high-redshift clusters presented in this work, we were able to obtain
optical identifications and to measure spectroscopic redshifts for a
significant number clusters at lower redshifts. These results will be
published in subsequent papers.

\acknowledgements

The work is supported by RSF grant 14-22-00271. In addition the
authors are grateful to TUBITAK, IKI, KFU and AST for the support of
the observations at RTT150 telescope. The work of IFB on the
organization of a part of the observations at RTT150 telescope was
partially funded by the subsidy 3.6714.2017/8.9 allocated to Kazan
Federal University for the state assignment in the sphere of
scientific activities. MVE is grateful to Basic Research Program P-7
of the Presidium of the Russian Academy of Sciences for partial
support of the observations at 1.6-m Sayan observatory telescope,
which is a part of ``Angara'' center.

\end{document}

%% file: fig_images_eng.tex
\begin{figure*}
  \centering
  PSZ2\,G$069.39\!+\!68.05$
  \medskip
  
  \includegraphics[width=0.98\columnwidth]{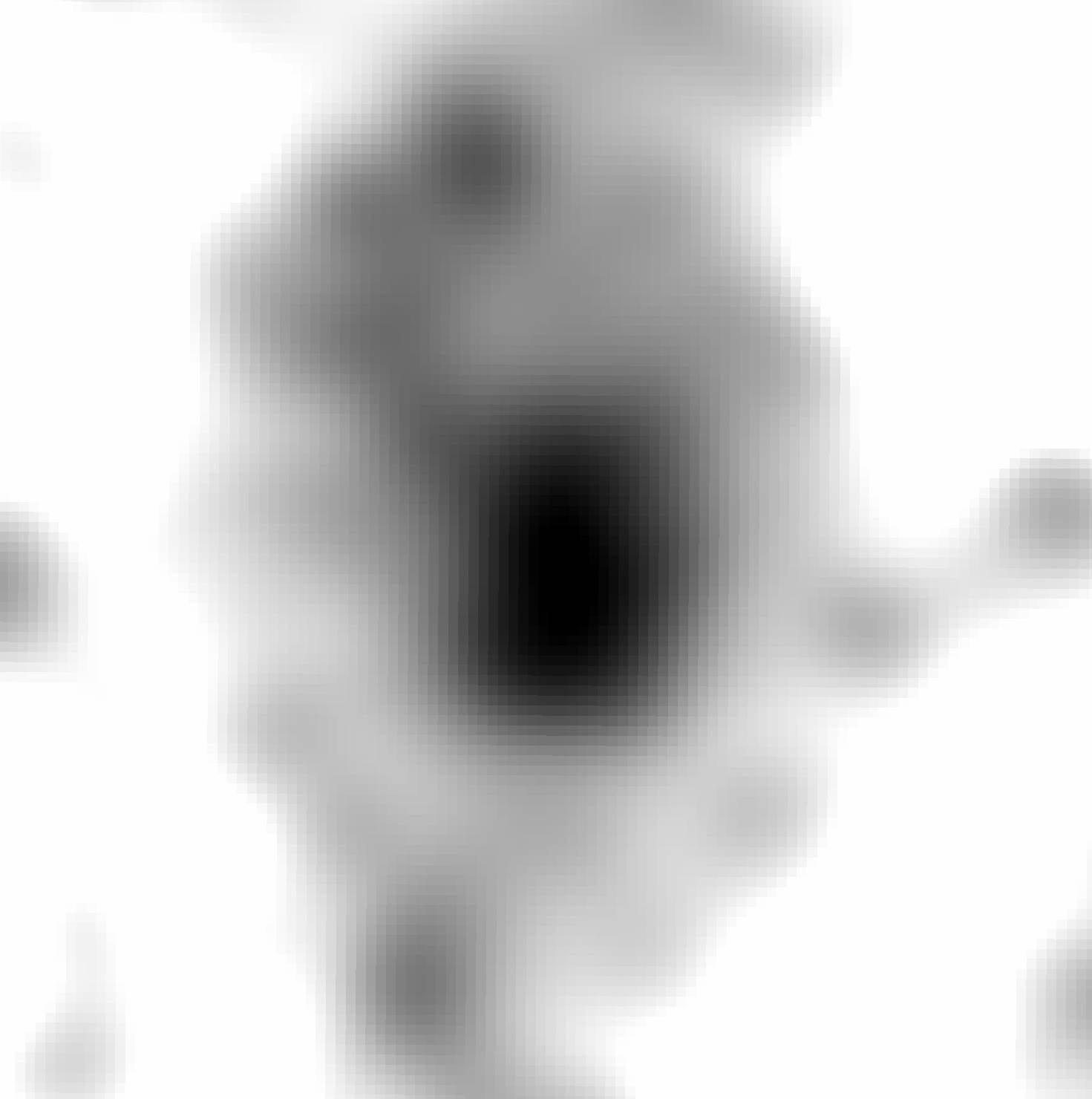} 
  ~
  \includegraphics[width=0.98\columnwidth]{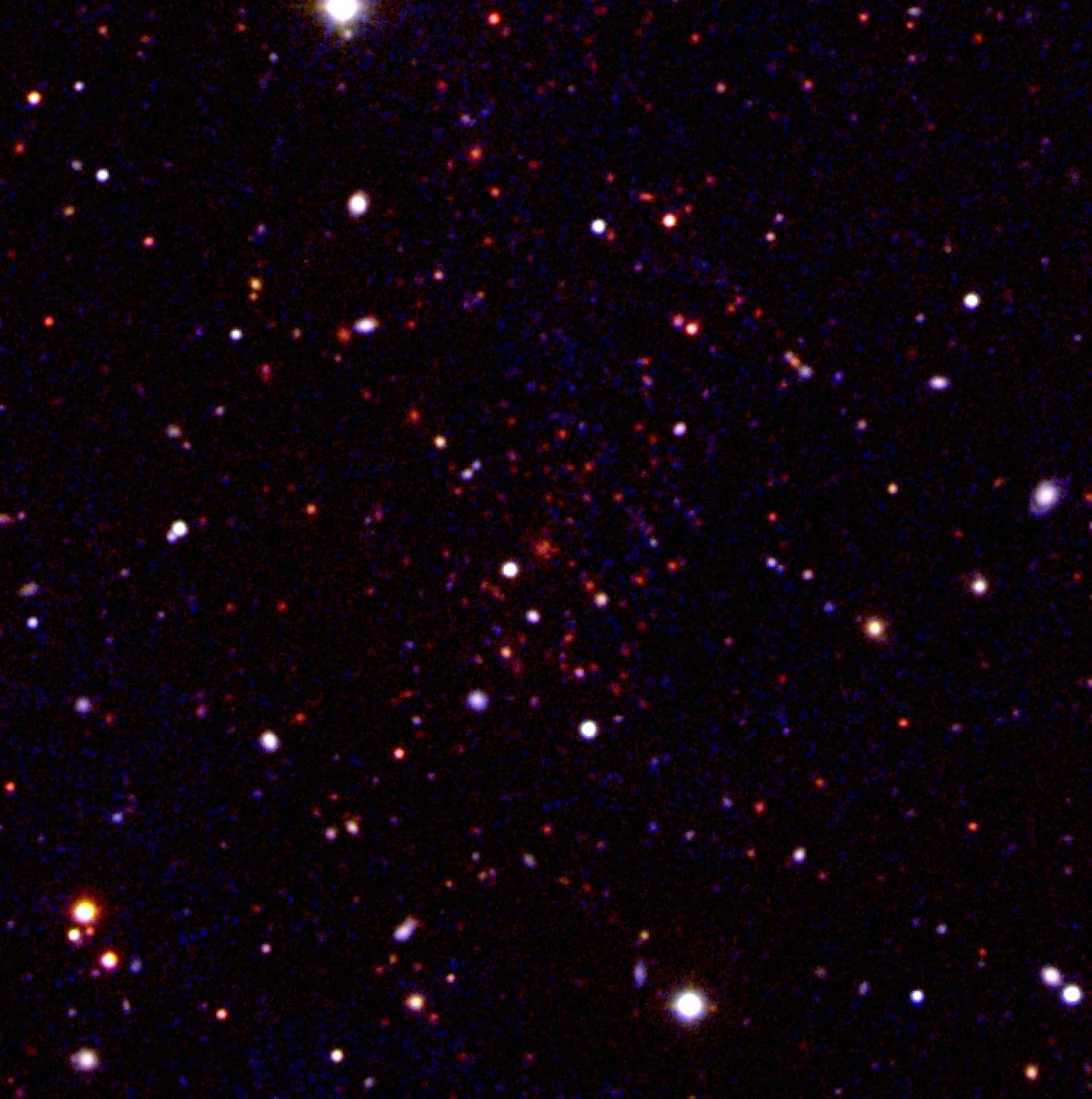}  

  \bigskip
  \bigskip

  PSZ2\,G$087.39\!-\!34.58$
  \medskip
  
  \includegraphics[width=0.98\columnwidth]{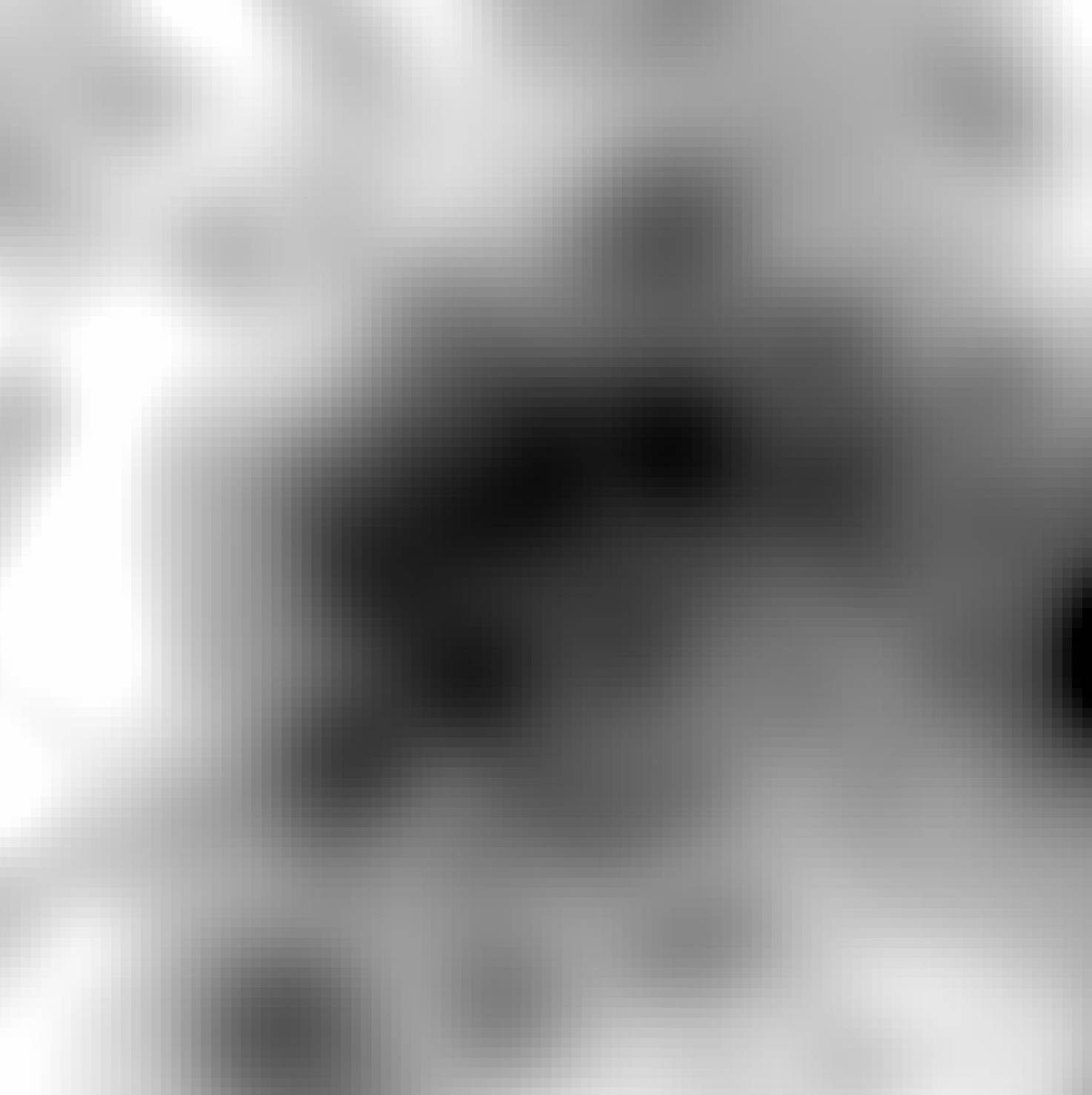} 
  ~
  \includegraphics[width=0.98\columnwidth]{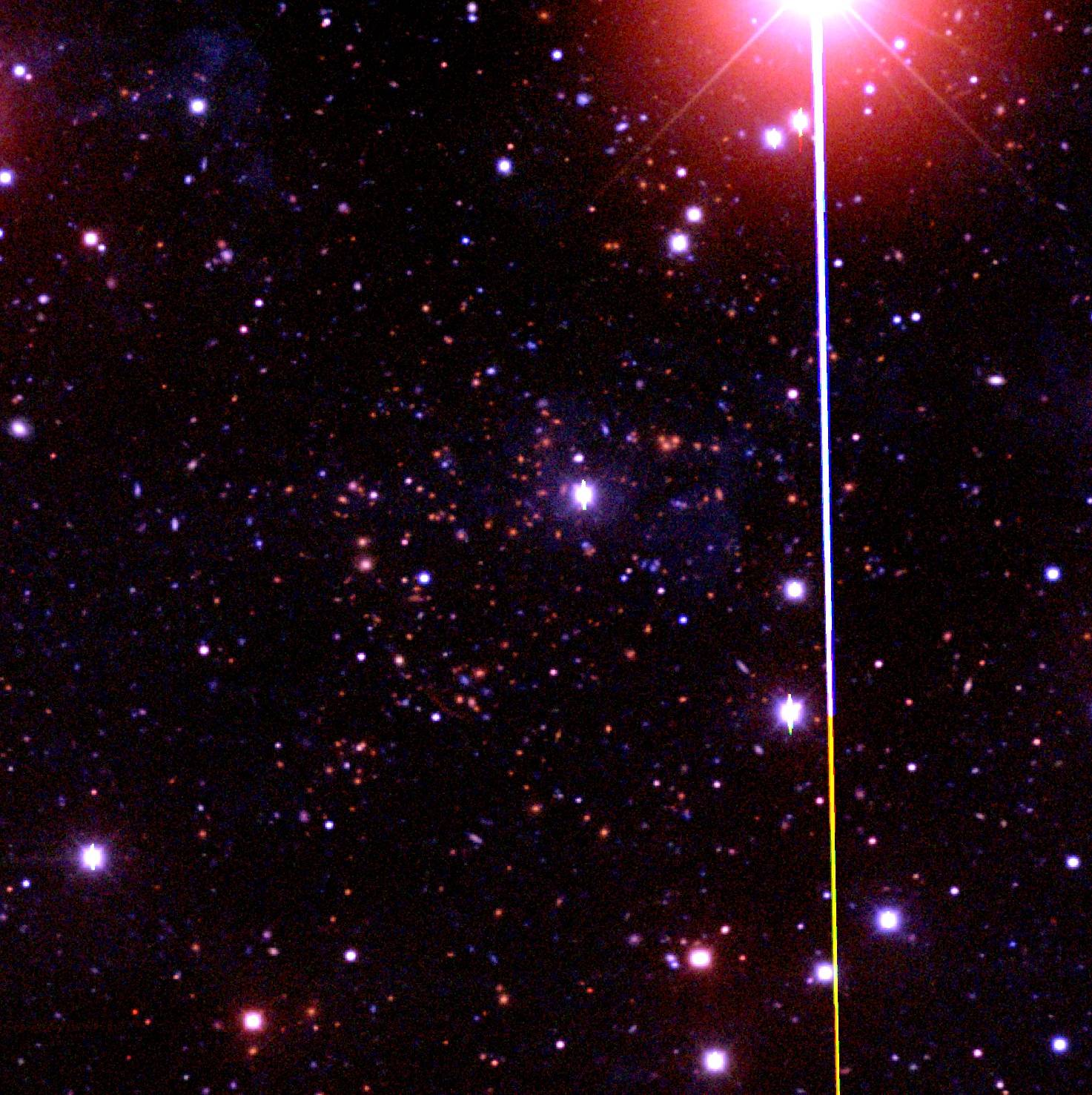}  
  \bigskip
  \bigskip
  \bigskip

  \caption{--- Left: the image in 3.4~$\mu$m band from WISE survey,
    cleaned from stars and convolved with $\beta$-model of $24\arcsec$
    radius. Right: pseudocolor image of the field in SDSS \emph{irg}
    filters (\emph{RGB}). The images are centered at the clusters
    optical centers, the image size is $5\arcmin \times 5\arcmin$.}
  \label{fig:fcharts}
\end{figure*}

\begin{figure*}
  \centering
  
  PSZ2\,G$092.69\!+\!59.92$
  \medskip
  
  \includegraphics[width=0.98\columnwidth]{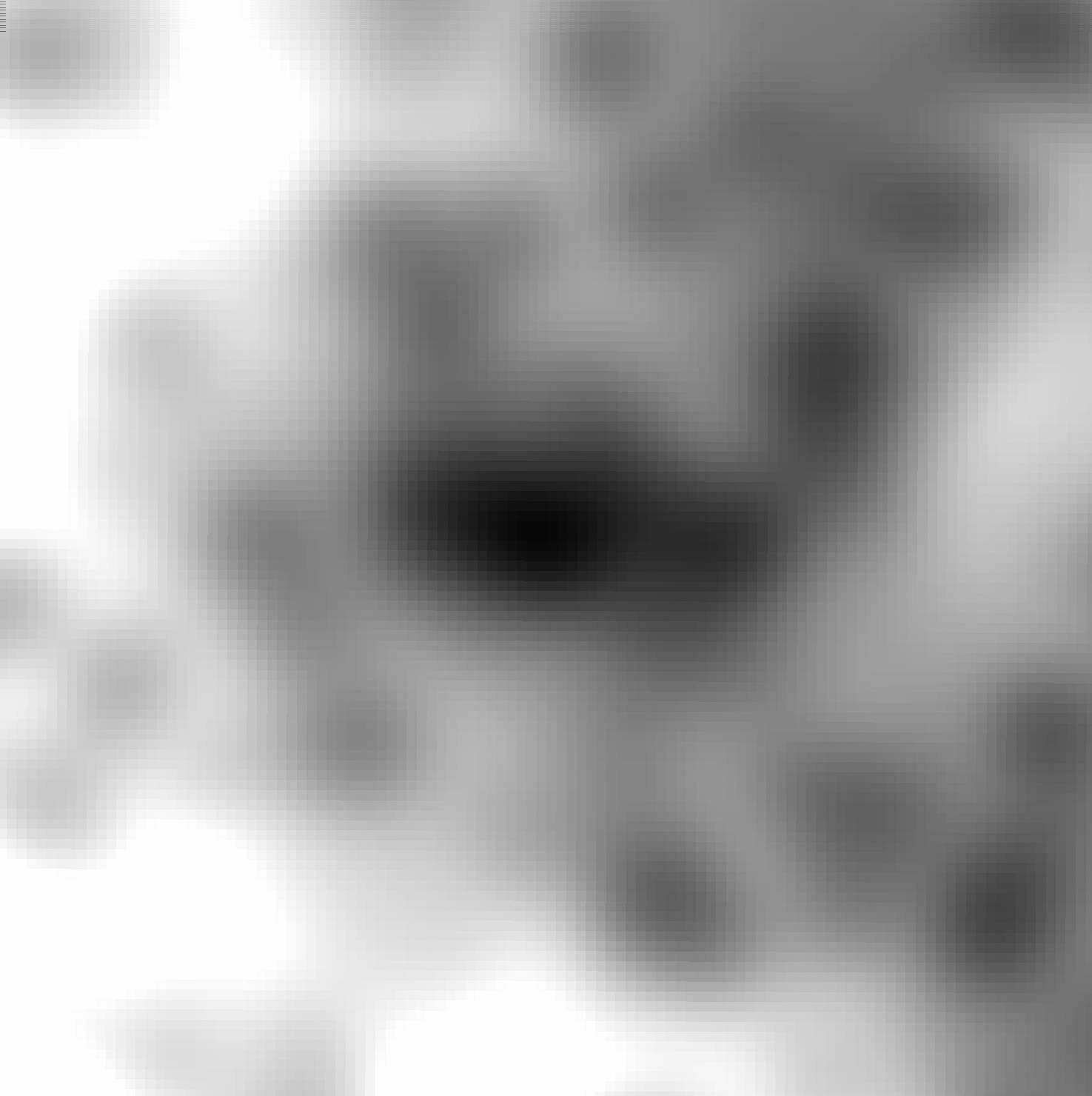} 
  ~
  \includegraphics[width=0.98\columnwidth]{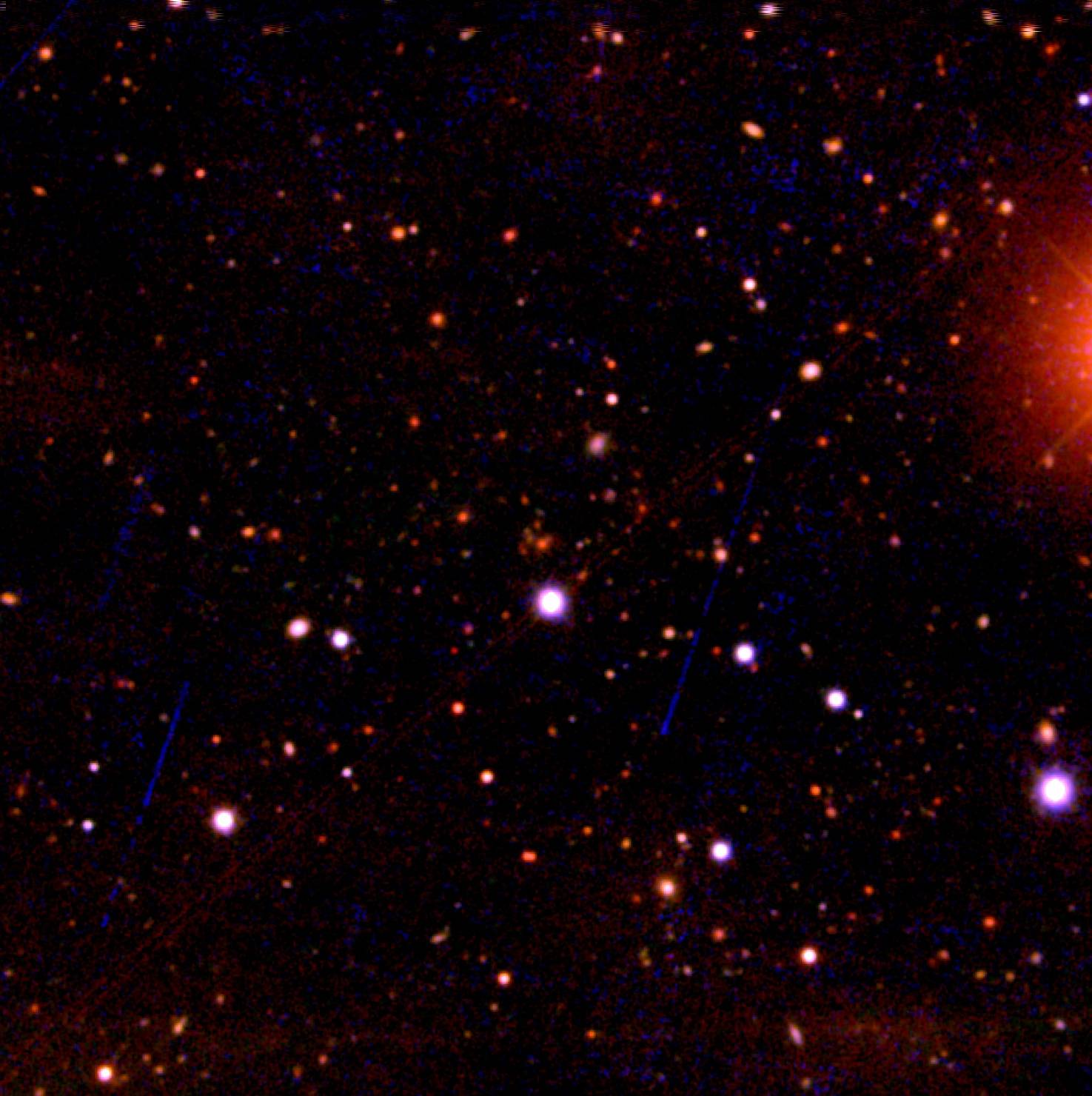} 

  \bigskip
  \bigskip
  
  PSZ2\,G$126.28\!+\!65.62$
  \medskip
  
  \includegraphics[width=0.98\columnwidth]{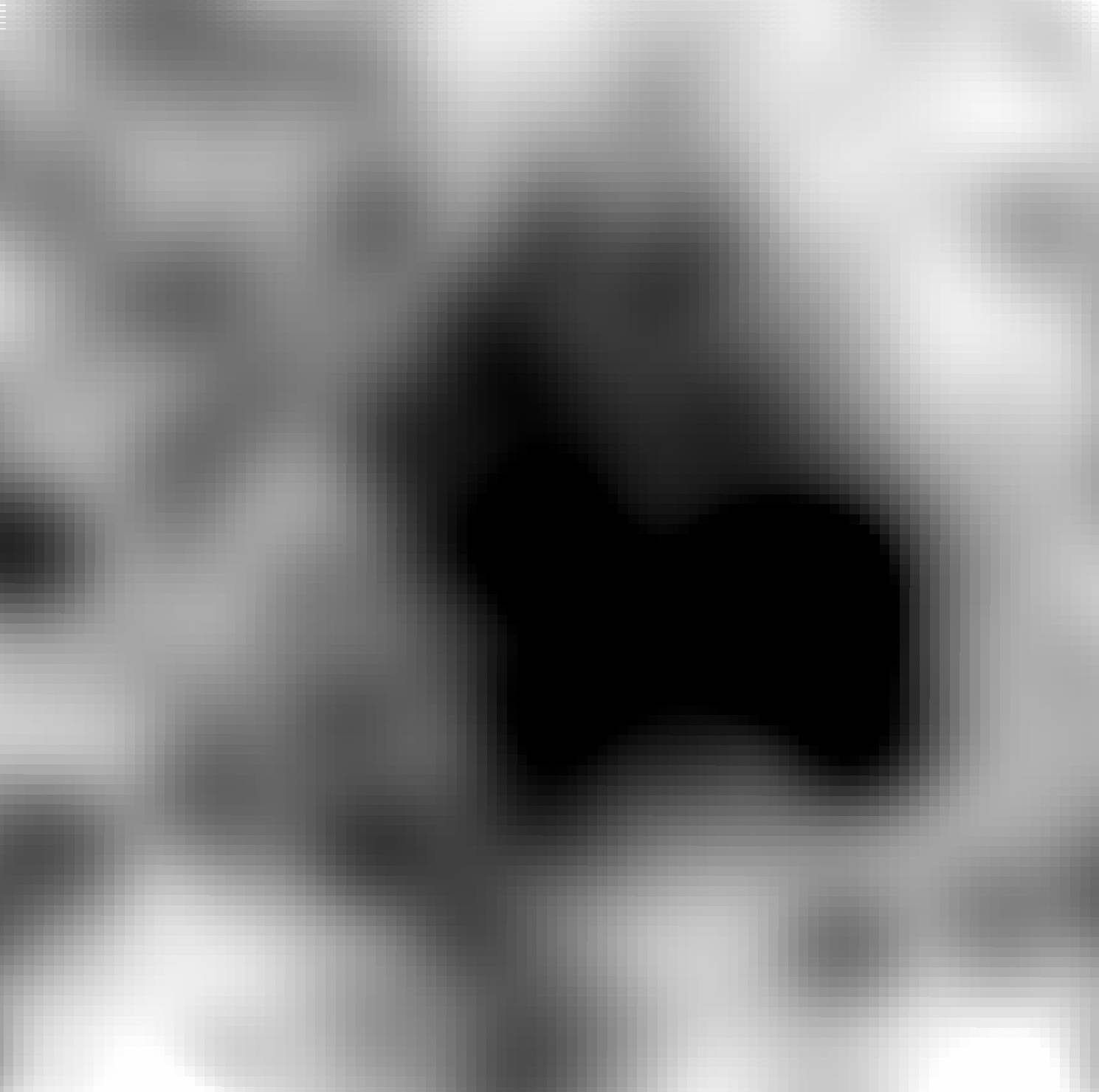} 
  ~
  \includegraphics[width=0.98\columnwidth]{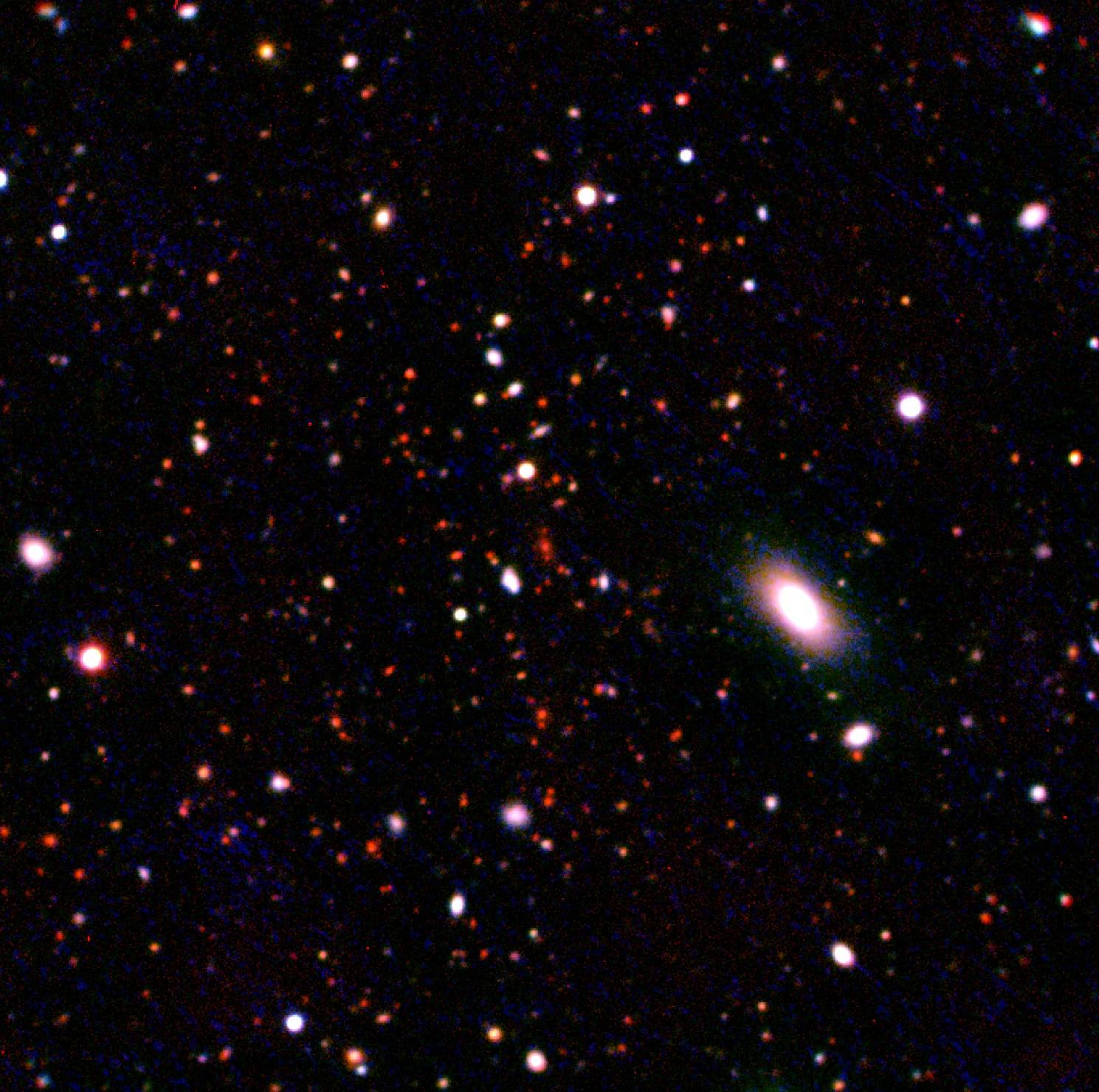}

  \bigskip
  \bigskip
  \bigskip

  {\small\textbf{Fig.~\ref{fig:fcharts}.} --- Continued.}
\end{figure*}

\begin{figure*}
  \centering

  PSZ2\,G$126.57\!+\!51.61$
  \medskip
  
  \includegraphics[width=0.98\columnwidth]{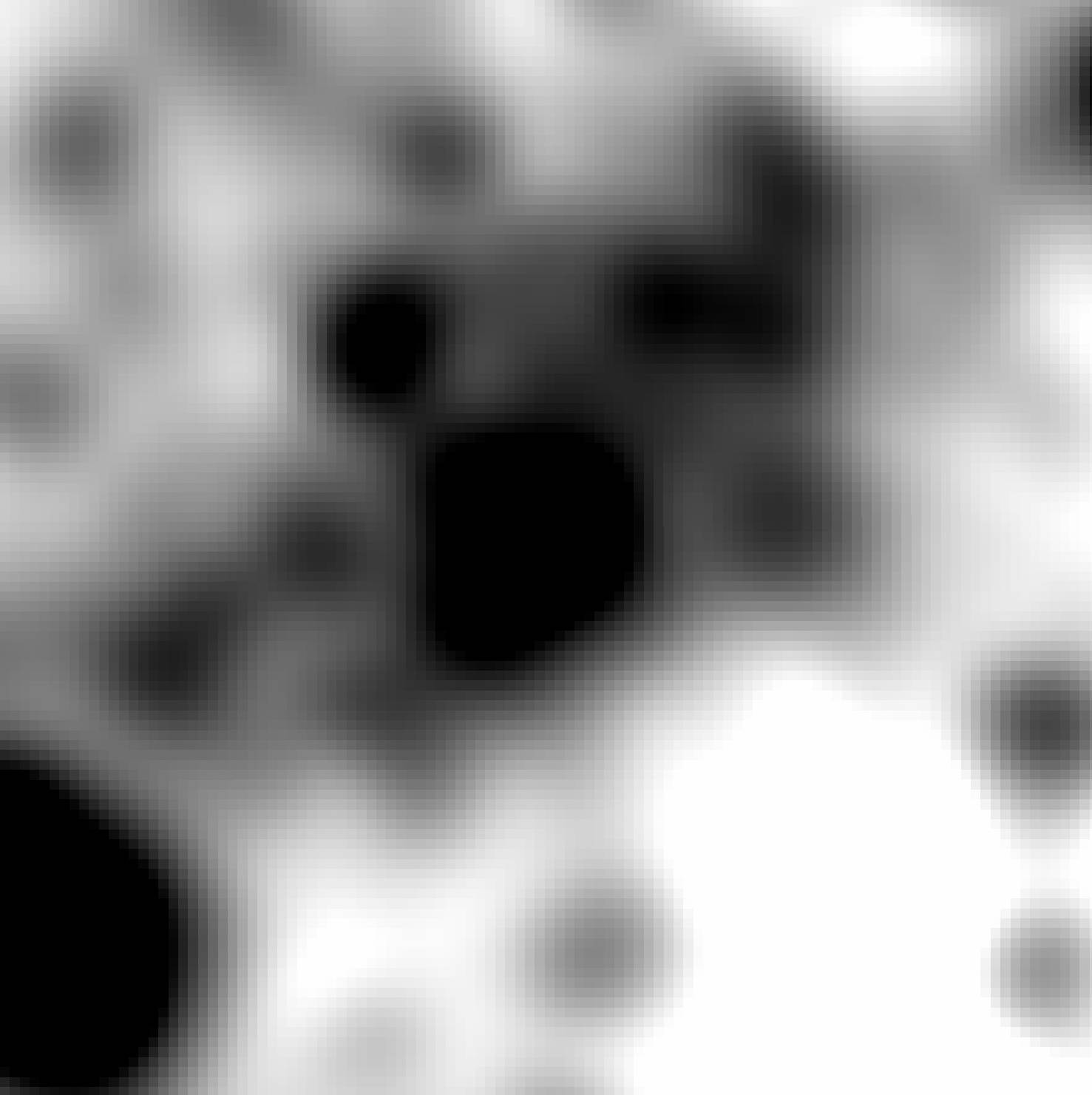} 
  ~
  \includegraphics[width=0.98\columnwidth]{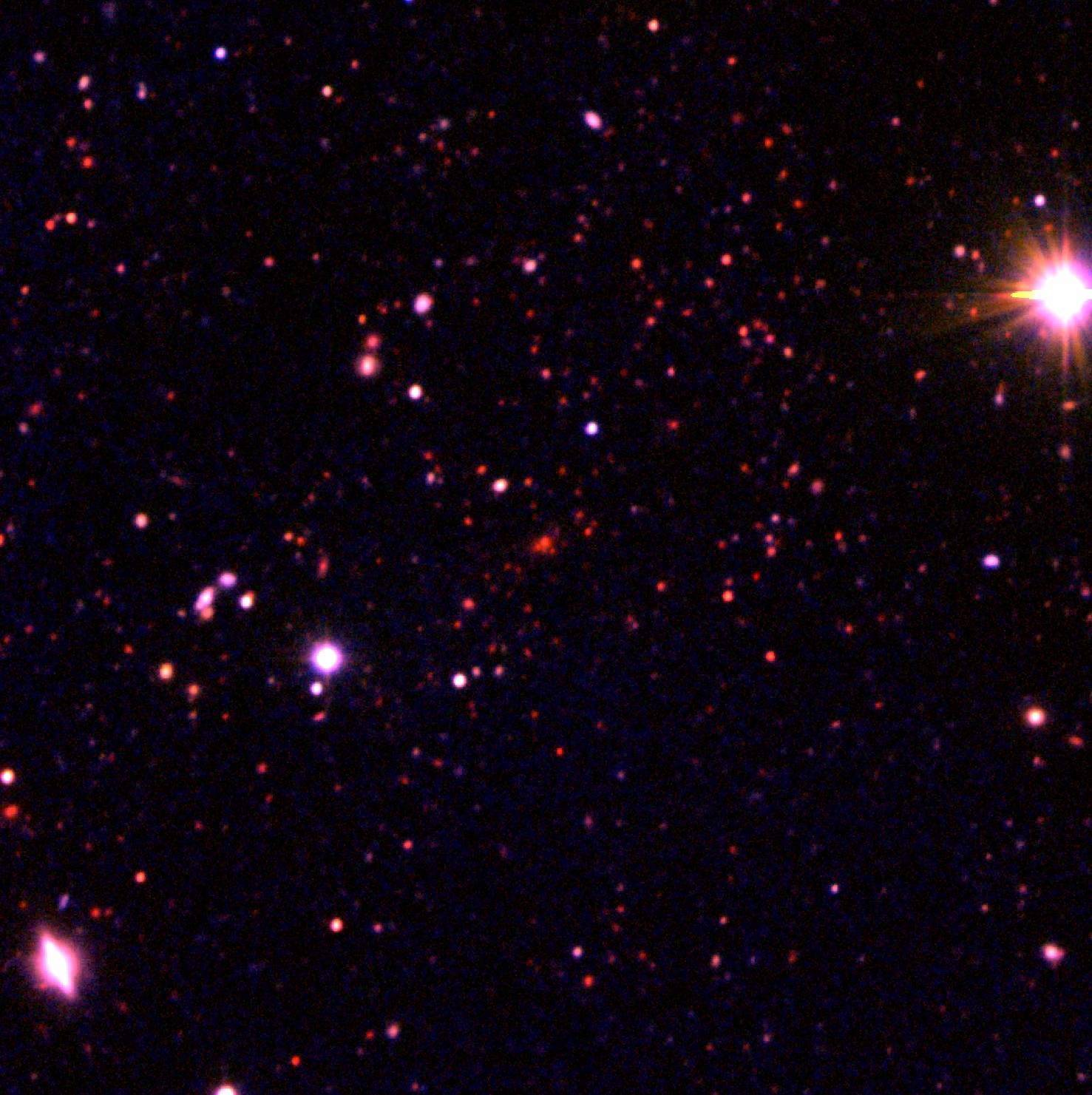} 

  \bigskip
  \bigskip

  PSZ2\,G$237.68\!+\!57.83$
  \medskip
  
  \includegraphics[width=0.98\columnwidth]{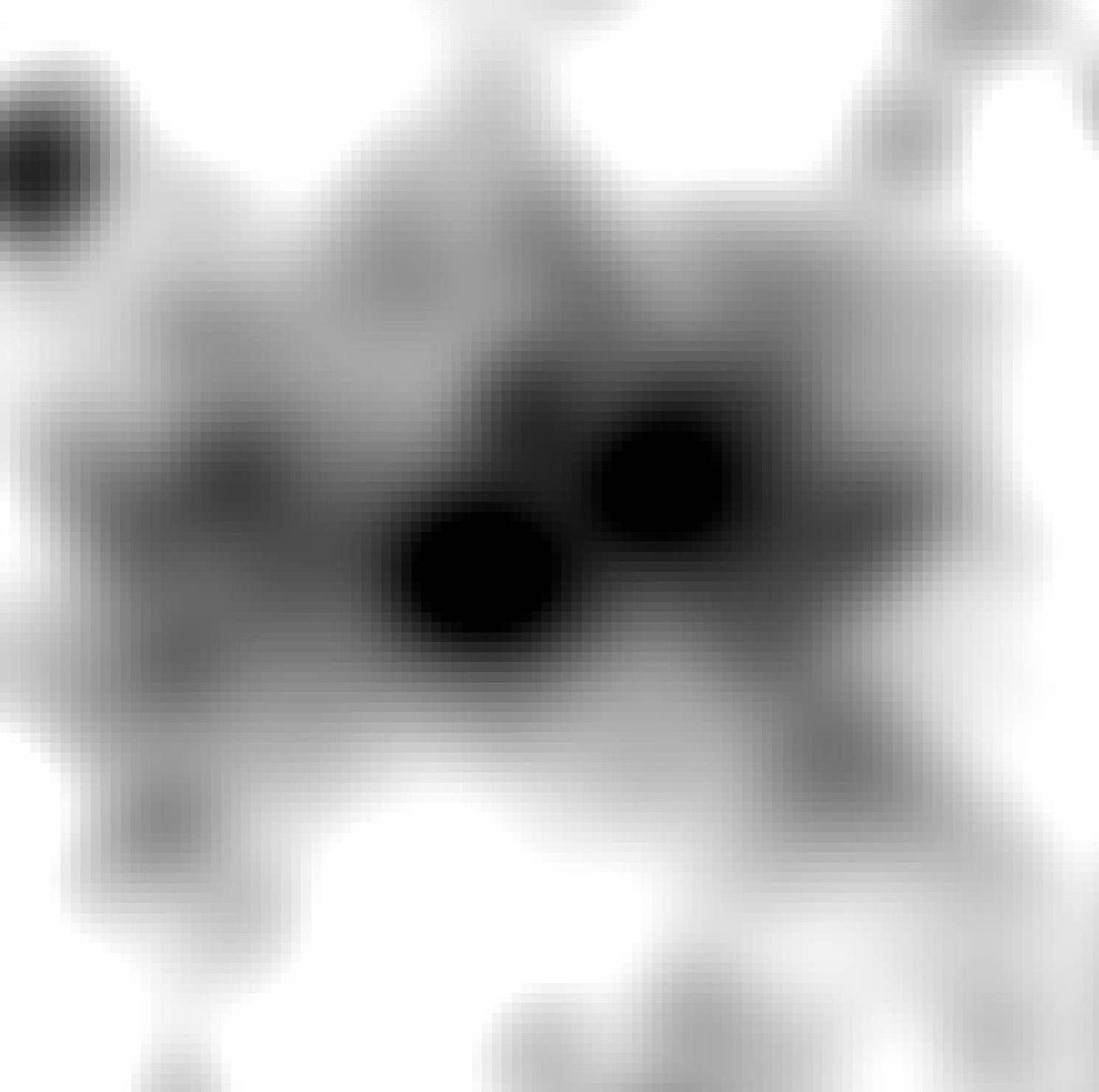} 
  ~
  \includegraphics[width=0.98\columnwidth]{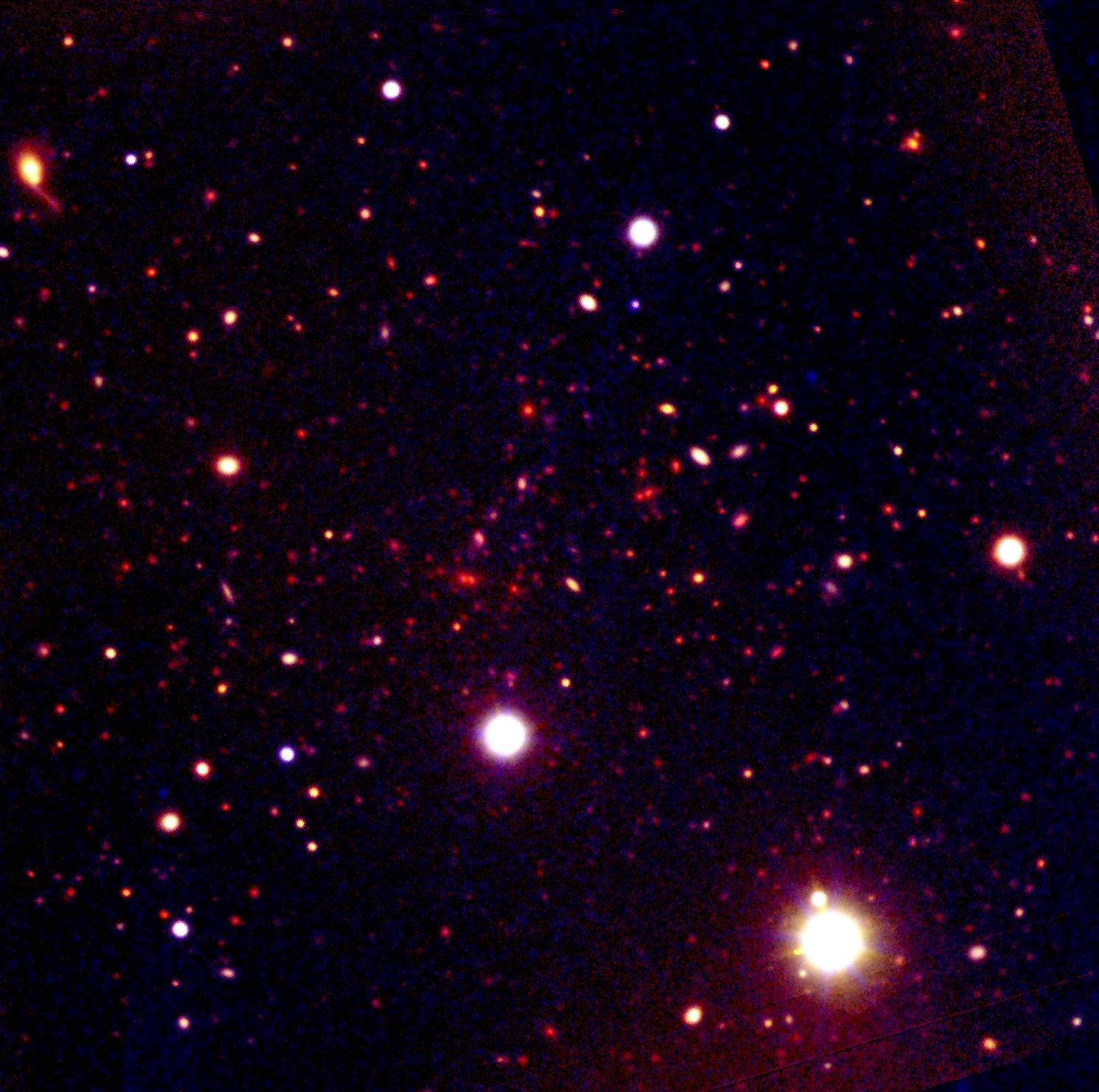}  

  \bigskip
  \bigskip
  \bigskip

  {\small\textbf{Fig.~\ref{fig:fcharts}.} --- Continued.}
\end{figure*}

\begin{figure*}
  \centering
  
  PSZ2\,G$343.46\!+\!52.65$
  \medskip
  
  \includegraphics[width=0.98\columnwidth]{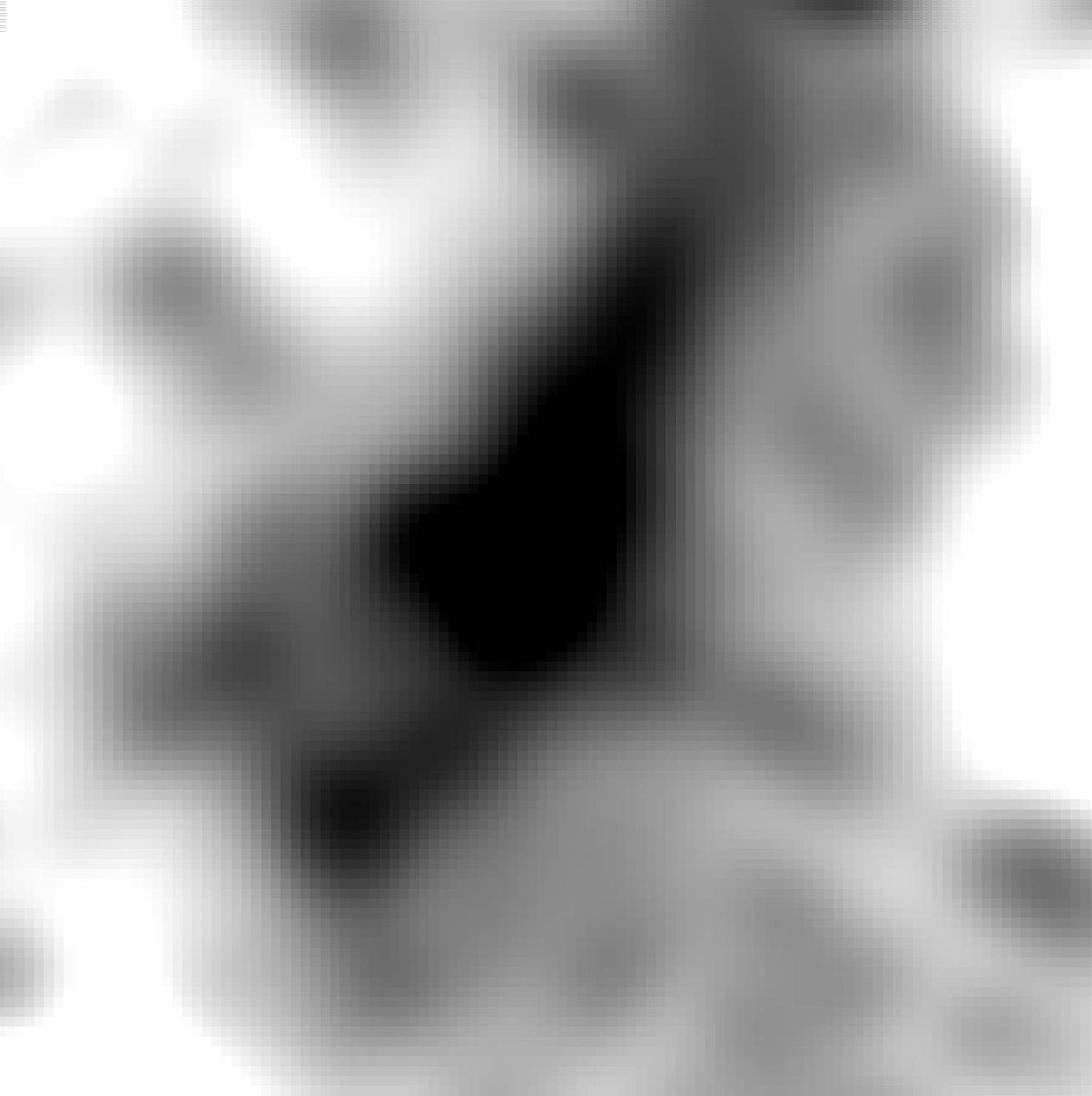} 
  ~
  \includegraphics[width=0.98\columnwidth]{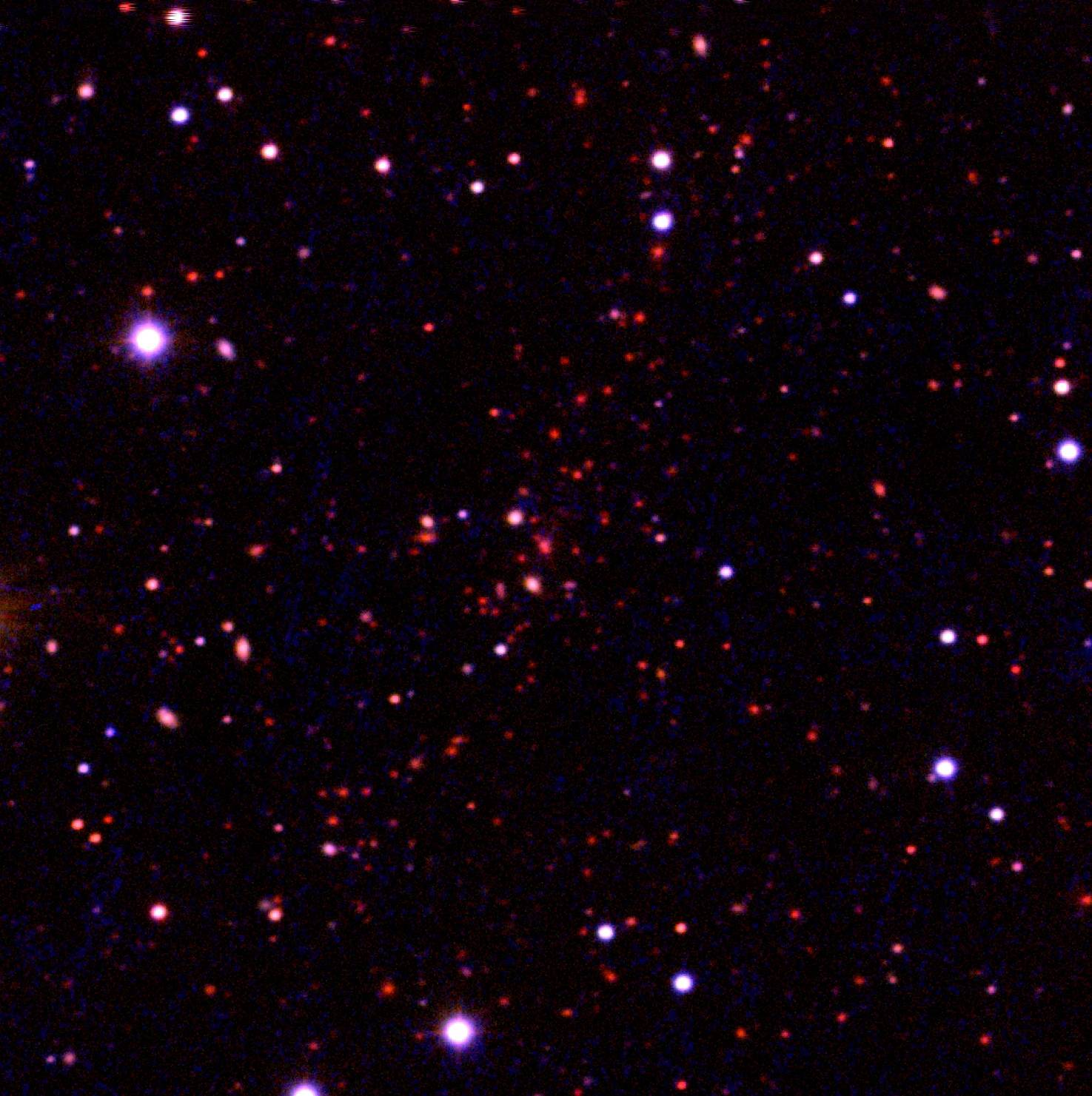} 

  \bigskip
  \bigskip

  {\small\textbf{Fig.~\ref{fig:fcharts}.} --- Continued.}
\end{figure*}

%% file: fig_spectra_eng.tex
\begin{figure*}
  \centering
  PSZ2\,G$069.39\!+\!68.05$
  
  \smfiguresmall{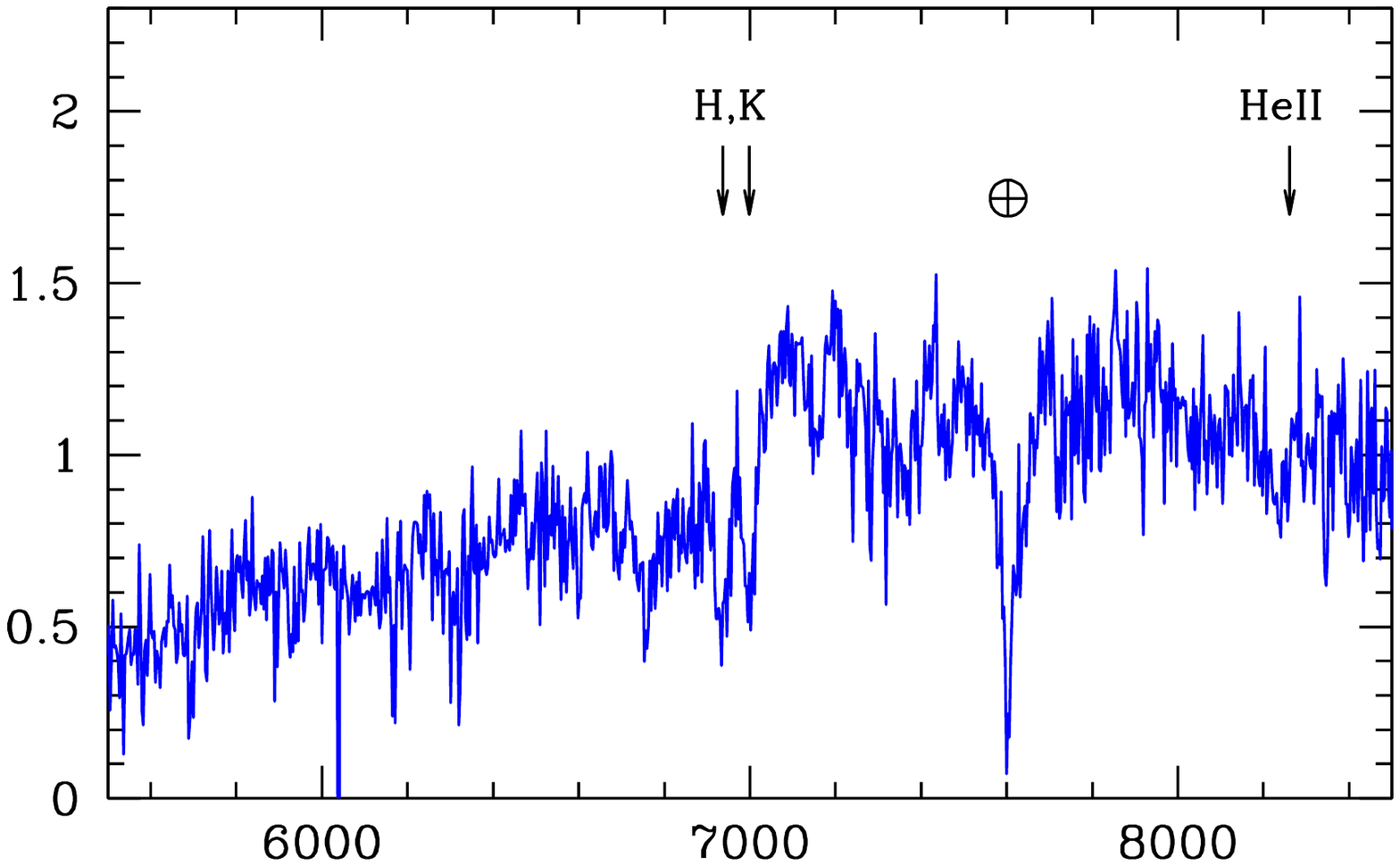}{$\lambda$, \AA}{Flux, $10^{-17}$~erg\,s$^{-1}$\,cm$^{-2}$}
  ~
  \smfiguresmall{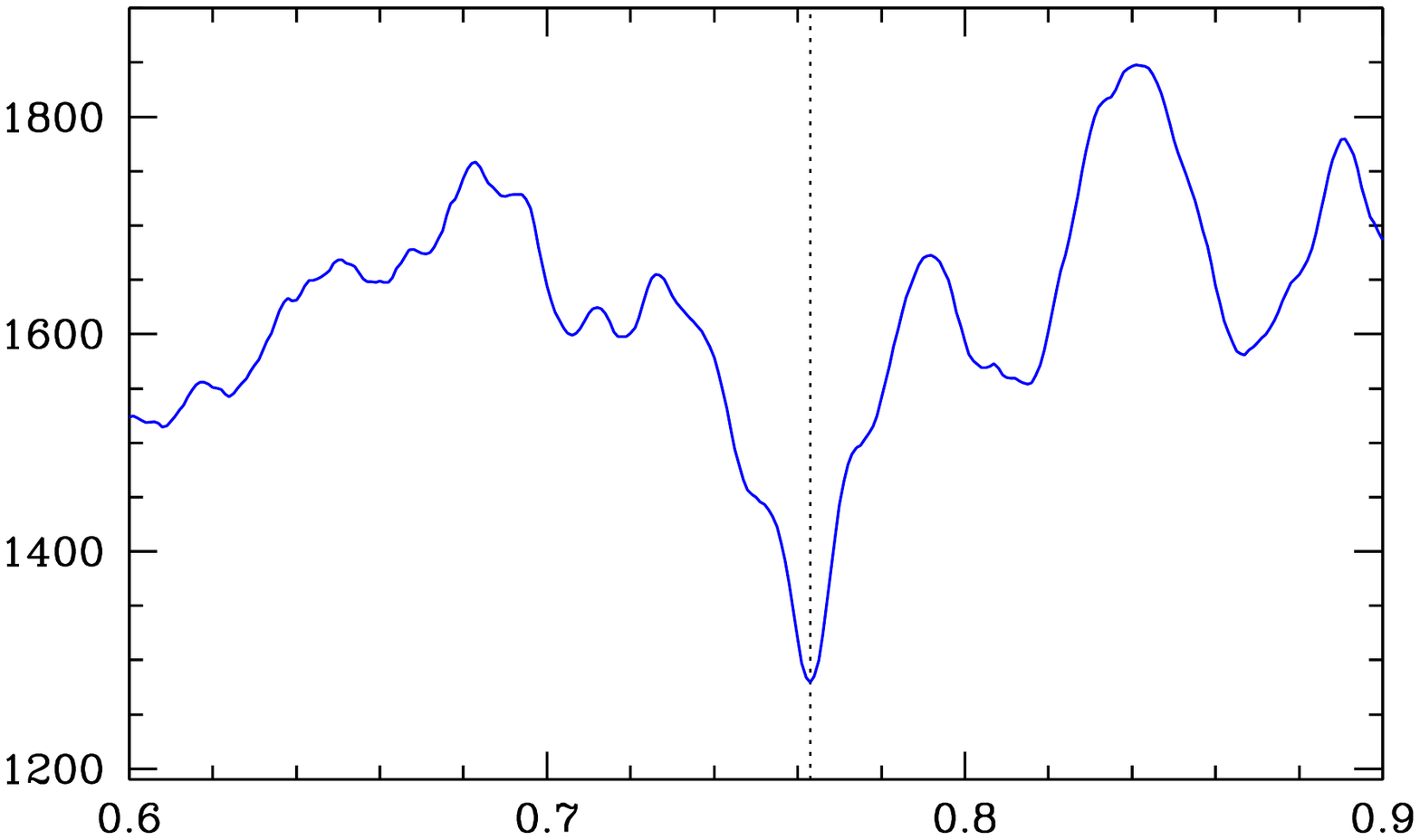}{$z$}{$\chi^2$} 

  \bigskip

  PSZ2\,G$092.69\!+\!59.92$
  
  \smfiguresmall{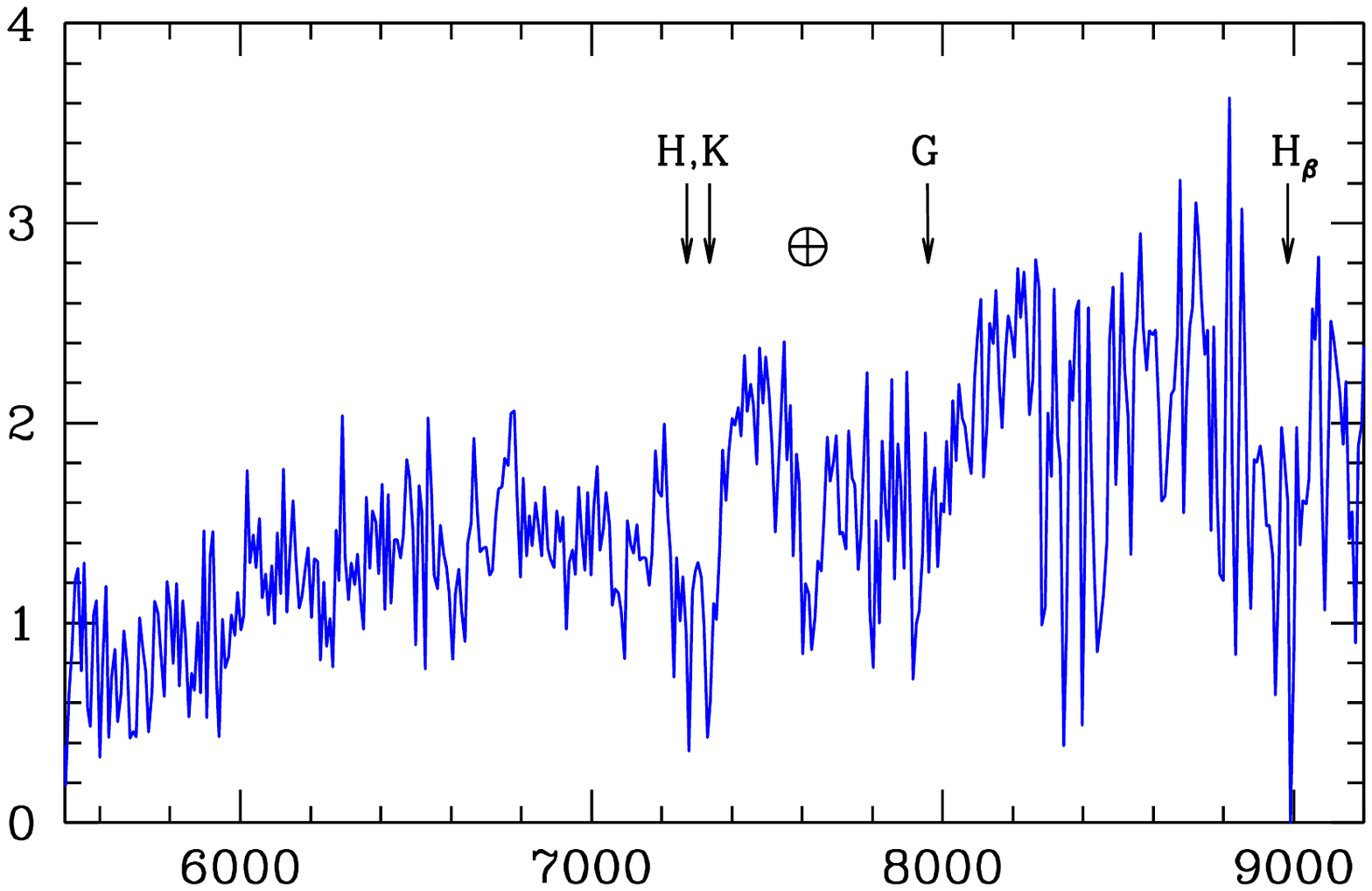}{$\lambda$, \AA}{Flux, $10^{-17}$~erg\,s$^{-1}$\,cm$^{-2}$}
  ~
  \smfiguresmall{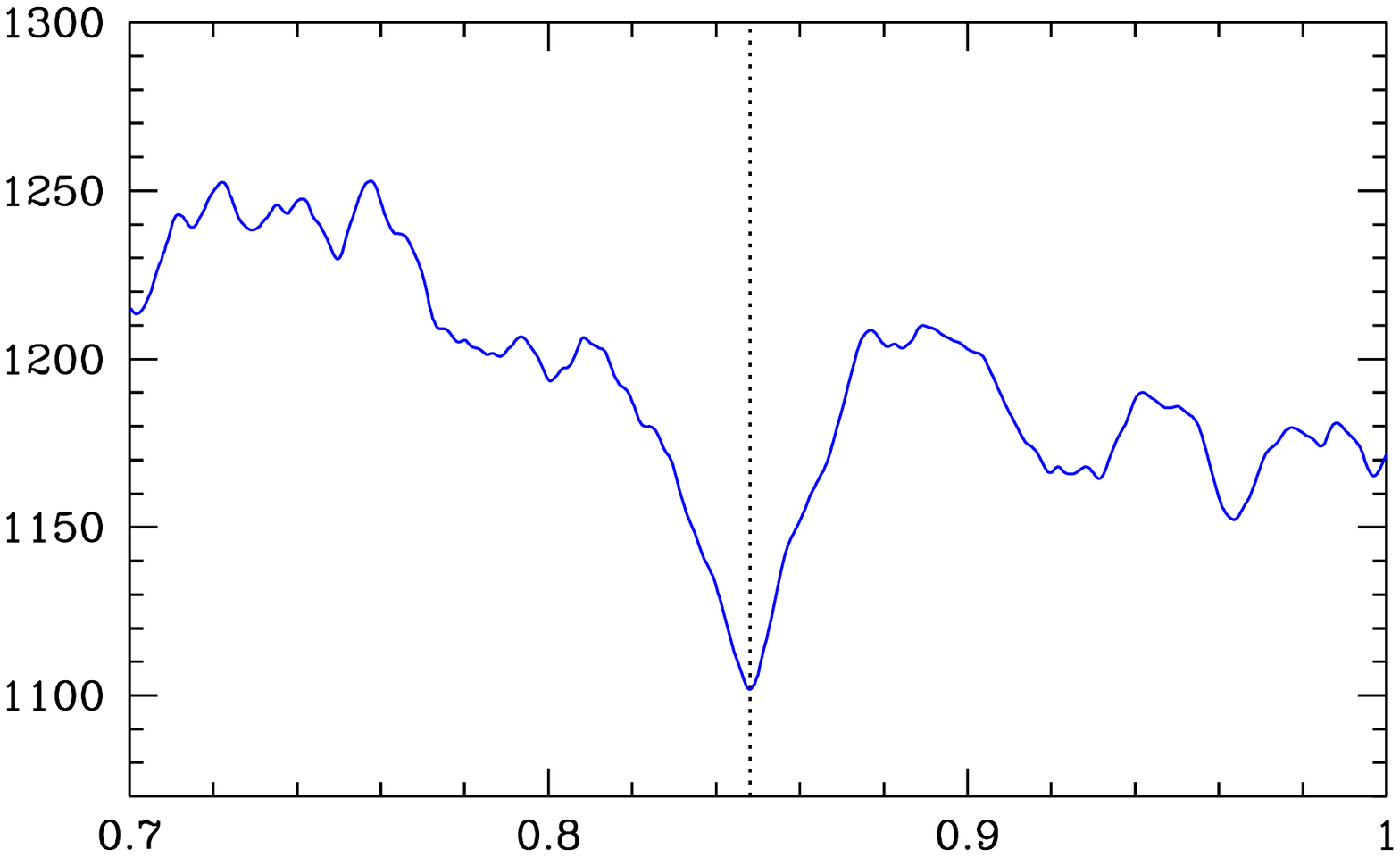}{$z$}{$\chi^2$} 
  
  \bigskip
  
  PSZ2\,G$126.28\!+\!65.62$
  
  \smfiguresmall{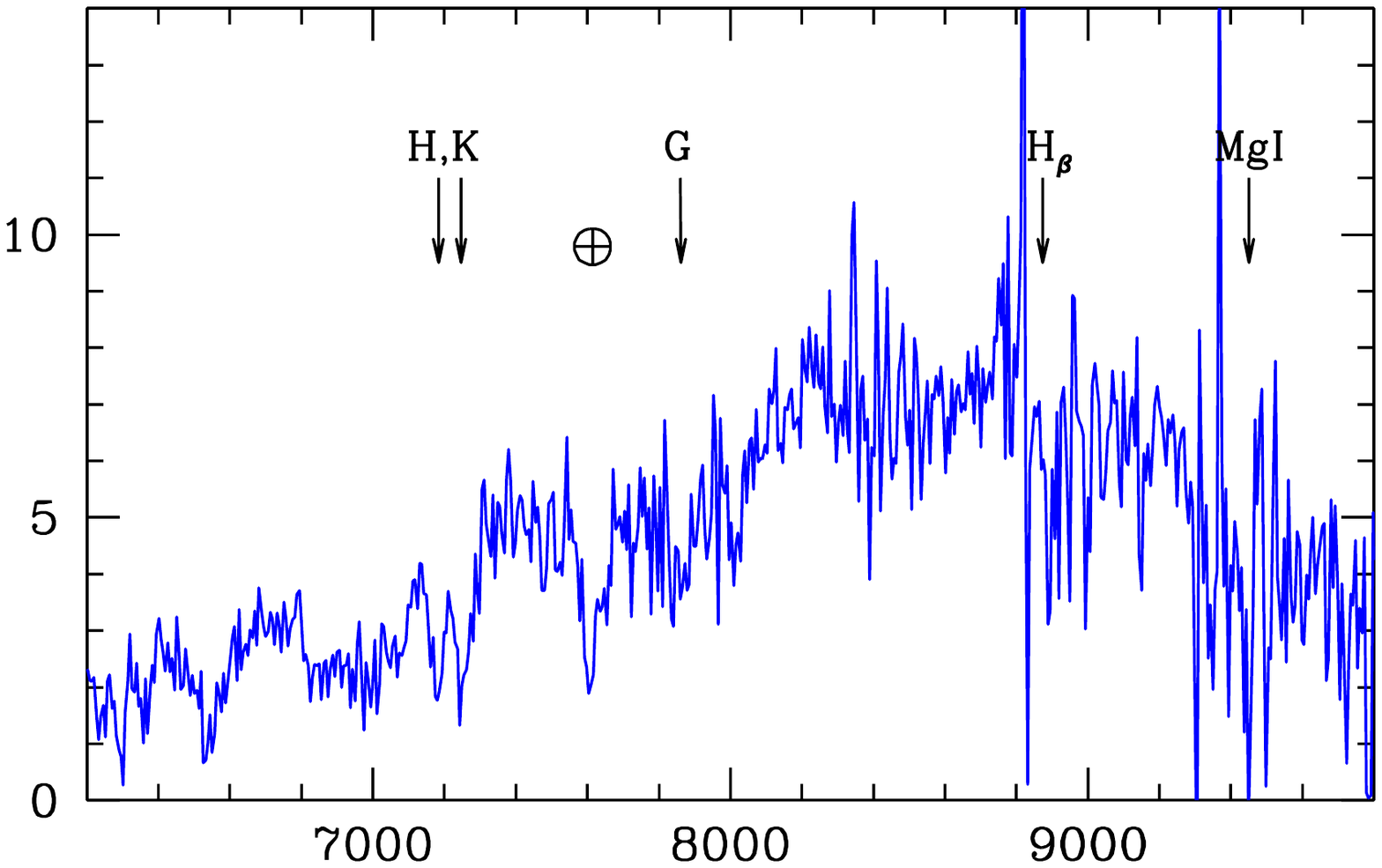}{$\lambda$, \AA}{Flux, $10^{-17}$~erg\,s$^{-1}$\,cm$^{-2}$}
  ~
  \smfiguresmall{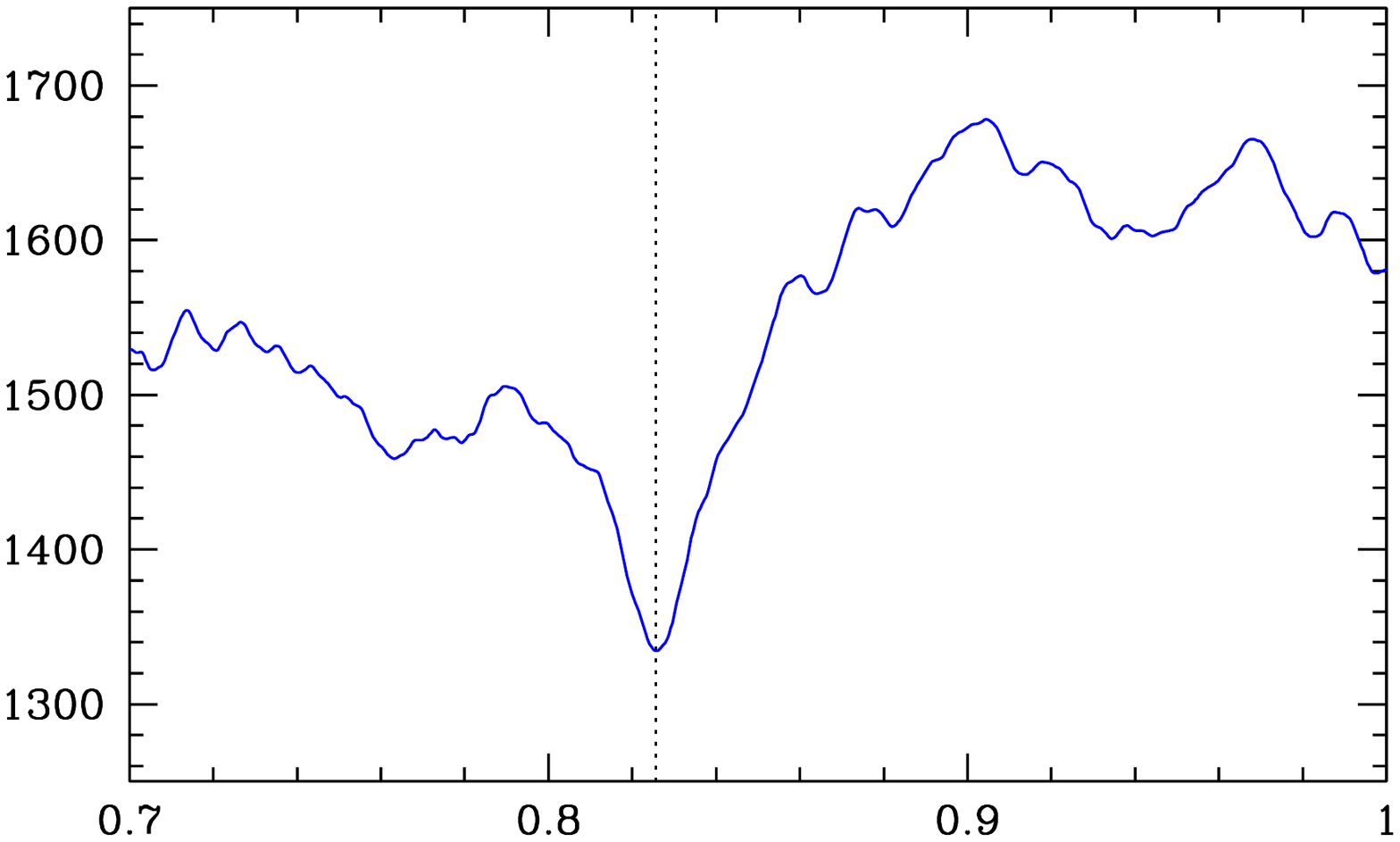}{$z$}{$\chi^2$} 
  
  \bigskip

  \caption{The examples of galaxy cluster redshift measurements. Left
    panel: the spectrum of brightest cluster galaxy with some spectral
    features indicated, obtained at BTA 6-m telescope using SCORPIO
    and SCORPIO-2 spectrographs. Right panel: the value of $\chi^2$
    obtained in result of cross correlation of this spectrum with the
    elliptical galaxy spectral template.}
  \label{fig:spec}
\end{figure*}

%% file: table_1_eng.tex
\begin{table*}
  \caption{High redshift galaxy clusters} 
  \label{tab:clres}
  \vskip 2mm
  \renewcommand{\arraystretch}{1.1}
  \renewcommand{\tabcolsep}{0.35cm}
  \centering
  \footnotesize
  \begin{tabular}{lcccll}
    \noalign{\doubleline}
    & \multispan2\hfil Coordinates (J2000)\hfil\\
    Name & $\alpha$ & $\delta$ & $z$ & $N_{gal}$ & Notes\\
    \noalign{\vskip 3pt\hrule\vskip 5pt}
    PSZ2\,G$069.39\!+\!68.05$ & $14~21~37.9$ & $+38~21~16$ &  $0.763$ & 3 & \emph{l},*\\ 
    PSZ2\,G$087.39\!-\!34.58$ & $22~49~09.7$ & $+19~43~57$ &  $0.771$ & 3 & \emph{l},* \\ 
    PSZ2\,G$092.69\!+\!59.92$ & $14~26~36.1$ & $+51~15~51$ &  $0.848$ & 2 &* \\ 
    PSZ2\,G$126.28\!+\!65.62$ & $12~42~23.4$ & $+51~26~22$ &  $0.820$ & 7 &\\ 
    PSZ2\,G$126.57\!+\!51.61$ & $12~29~47.8$ & $+65~21~13$ &  $0.815$ & 1 &\emph{c} \\ 
    PSZ2\,G$237.68\!+\!57.83$ & $10~53~16.3$ & $+10~52~46$ &  $0.892$ & 1 &* \\ 
    PSZ2\,G$343.46\!+\!52.65$ & $14~24~23.1$ & $-02~43~48$ &  $0.713$ & 3 &\\ 
    \noalign{\vskip 3pt\hrule\vskip 5pt}
  \end{tabular}
  
  \begin{minipage}{0.78\linewidth}
    Notes:
    \smallskip
    
    
    \emph{l} --- gravitational lensing arc is found; 

    * --- discussed in text; 

    \emph{c} --- the cluster from cosmological sample of \emph{Planck} SZ survey; 

  \end{minipage}

\end{table*}